\documentclass[]{aastex631}

\usepackage{amsmath}
\usepackage{float}

\defcitealias{2024ApJ...966...68H}{AH24}
\defcitealias{2025ApJ...991..208Z}{SZ25}
\usepackage{multirow}
\begin{document}

\title{Signatures of Damping Nonlinear Oscillations by KHI-induced Turbulence in Synthetic Observations}
%\title{Nonlinear Damping of Large-Amplitude Transverse Loop Oscillation: KHI-induced Turbulence and Observational Signatures}

\correspondingauthor{Sihui Zhong}
\email{sihui.zhong@kuleuven.be}

\author[0000-0002-5606-0411]{Sihui Zhong}
\affiliation{Centre for mathematical Plasma Astrophysics, Department of Mathematics, KU Leuven, Leuven, BE-3001, Belgium}
\affiliation{Engineering Research Institute \lq\lq Ventspils International Radio Astronomy Centre (VIRAC)\rq\rq\ of Ventspils University of Applied Sciences,\\ Inzenieru iela 101, Ventspils, LV-3601, Latvia}

\author[0000-0002-0851-5362]{Andrew Hillier}
\affiliation{Department of Mathematics and Statistics, University of Exeter,
Exeter, EX4 4QF, UK}

\author[0000-0002-7008-7661]{I\~nigo Arregui}
\affiliation{Instituto de Astrof\'{\i}sica de Canarias, V\'ia L\'actea S/N, E-38205 Laguna, Tenerife, Spain}
\affiliation{Departmento de Astrof\'{\i}sica, Universidad de La Laguna, E-38206 Laguna, Tenerife, Spain}

\begin{abstract}
Large-amplitude decaying kink oscillations of coronal loops are strongly influenced by nonlinear processes, such as Kelvin–Helmholtz instability (KHI) and turbulence, though comprehensive theory and observational confirmation remain limited. Building on the recently developed theory on nonlinear damping by KHI-induced turbulence in impulsively driven transverse loop oscillations, %we investigate the signatures of nonlinear damping using 3D magnetohydrodynamic simulations and synthetic Extreme Ultraviolet images via forward modelling. 
we investigate its observational signatures using 3D magnetohydrodynamic simulations and forward-modelled EUV images.
The simulated oscillations exhibit time-varying frequency shifts and damping rates, which are broadly consistent with nonlinear turbulence-damping theory. Additionally, they exhibit excitation of higher-order modes, slightly increased periods relative to the linear kink period, and reduced displacement amplitudes. These features are generally preserved in synthetic observations, though resolving higher-order modes requires higher spatial resolution than currently available. 
For loops embedded in a hotter background, hotter channels (e.g., 193\,\AA) are more sensitive to boundary dynamics, thus their oscillations decay faster with smaller displacements and larger phase shifts than those in cooler channels (e.g., 171\,\AA).
Comparisons of simulated and synthetic oscillations show close agreement at the early stage. At later times, synthetic oscillations exhibit smaller displacements and larger phase shifts, due to turbulence-induced asymmetry in the loop cross-section. %[Reproducing this enhanced late-stage decay while preserving the early-time behaviour requires a slightly higher density contrast and a greater fraction of the boundary region participating in collective oscillation in the model.] 
Bayesian fitting shows that the initial oscillation amplitude and kink period are robustly constrained, whereas parameters controlling the damping profile are degenerate, indicating that additional observables would aid reliable seismological inference. These results provide a quantitative basis for identifying nonlinear damping and detecting KHI-driven turbulence in transverse loop oscillations. %(234/250)
%In general, loop oscillations in 171\,\AA\ trace core motions, while hotter channels depict the loop boundary. 

\end{abstract}



\section{Introduction} \label{sec:intro}

Transverse oscillations of coronal loops are a prominent manifestation of magnetohydrodynamic (MHD) waves in the solar corona, providing powerful diagnostics of plasma properties through MHD seismology \citep{2020ARA&A..58..441N,2024RvMPP...8...19N}.
Large-amplitude transverse oscillations of coronal loops, first observed in 1998 by the \textit{Transition Region and Coronal Explorer} (TRACE) \citep{1999ApJ...520..880A,1999Sci...285..862N}, are interpreted as standing kink waves in magnetic cylinders, as theoretically formulated in the 1970s-1980s \citep{1975IGAFS..37....3Z,1983SoPh...88..179E}. These decaying kink oscillations are typically excited by impulsive coronal eruptions \citep{2015A&A...577A...4Z}, exhibit rapid damping, and generally appear as a standing mode \citep{2019ApJS..241...31N}. 
While kink waves have been suggested as a possible contributor to coronal heating \citep{2020SSRv..216..140V}, transforming all the energy associated with a typical resonantly damped kink oscillation into internal energy can only raise the plasma temperature by $\sim10^5$~K  \citep{2018RNAAS...2..196T}, which is insufficient to balance the coronal radiative loss.
%Moreover, the interplay between observed wave properties and MHD wave theory allows plasma diagnostics, forming the basis of MHD seismology \citep{2024RvMPP...8...19N}.

Large-amplitude decaying kink oscillations are considered to be nonlinear, due to their amplitudes being greater than the loop radius and their amplitude-dependent strong damping \citep{2019ApJS..241...31N}. Large-amplitude perturbations can nonlinearly generate additional modes, such as sausage ($m=0$) and fluting ($m=2$) modes, both of which oscillate at twice the kink mode ($m=1$) frequency \citep{2017SoPh..292..111R}, as well as higher-order ($m>2$) harmonics \citep[e.g.,][]{2018ApJ...853...35T}. Those modes may enhance damping through resonant coupling with the fundamental kink mode \citep{2010PhPl...17h2108R} or via nonlinear cascade to smaller scales \citep{2021ApJ...910...58V}. %{However, recent simulations show that the $m\geq2$ modes are overdamped by the resonant absorption and flow instabilities \citep{2025A&A...693A.201S}.} 
Analytical solutions for nonlinear kink oscillations in stratified loops predict frequency drifts proportional to the square of the oscillation amplitude \citep{2014SoPh..289.1999R}, and additional effects such as ponderomotive forces at the loop apex \citep{2004ApJ...610..523T}. %Additionally, nonlinearity may produce pondermotive force \citep{2004ApJ...610..523T} at the loop apex.

Among nonlinear processes, the Kelvin-Helmholtz instability (KHI) is particularly important because it can alter both the damping and energy dissipation of kink oscillations. First identified in 3D high-resolution MHD simulations \citep{2008ApJ...687L.115T}, KHI develops in oscillating loops due to azimuthal velocity shear between the loop boundary and its surrounding stationary ambient plasma \citep{1983A&A...117..220H}, and can occur even at modest velocity amplitudes \citep{2014ApJ...787L..22A}. Its onset is accelerated by steeper transverse density gradients and larger oscillation amplitudes \citep{2008ApJ...687L.115T,2016A&A...595A..81M}.
%and its induced turbulence, first revealed in 3D high-resolution MHD simulations \citep{2008ApJ...687L.115T}. In an oscillating loop, KHI is triggered by the azimuthal velocity shear between the loop and its stationary ambient \citep{1983A&A...117..220H}, not necessarily with a large velocity amplitude \citep{2014ApJ...787L..22A}. The steeper the transition layer of the oscillating loop, the greater the driving velocity, the quicker KHI sets in \citep{2008ApJ...687L.115T,2016A&A...595A..81M}.
Morphologically, KHI manifests as small-scale roll-ups and vortices around the loop boundary \citep[e.g.,][]{2008ApJ...687L.115T,2016ApJ...830L..22A}. In spectroscopic observations, KHI produces an arrow-shaped Doppler velocity shift oscillation with half the kink period and line broadening near the boundary \citep{2017ApJ...836..219A}. \citet{2017ApJ...836..219A} estimate that detecting these features requires a spatial resolution of $0.33$\arcsec and a spectral resolution of $25$~km~s$^{-1}$.
The significance of KHI extends beyond generating fine structures, as it directly influences damping efficiency and energy cascade to small scales.
In the linear regime, the damping of kink oscillation is dominated by resonant absorption \citep{2002A&A...394L..39G,2002ApJ...577..475R}. When KHI comes into play, it accelerates the decay by transferring the energy to small-scale structures, with the rate increasing with the driving velocity \citep{2016A&A...595A..81M}. 
The heating generated by KHI and its induced turbulence largely depends on its input. Studies in which the impulsive energy input is of weak to moderate magnitude report negligible heating \citep[e.g.,][]{2016A&A...595A..81M,2017A&A...607A..77H,2024ApJ...966...68H}. %main role in mixing, leading to turbulence

%[squashing effect due to the inertia of dense plasma.
%higher-order modes have higher power at the beginning \citep{2019FrP.....7...85A}.]

Direct observational evidence of nonlinearity in the solar corona, especially KHI and turbulence, remains elusive. This challenges both the validation of nonlinear wave theory and the applicability of MHD seismology derived from either linear or nonlinear frameworks. Most previous numerical studies of KHI have emphasised fine-scale features that are below the resolution limits of current instrumentation. Given that KHI efficiently mixes plasma, it can even destroy an oscillating loop composed of multiple strands, rendering the structure fully turbulent \citep{2016ApJ...823...82M}. %In such a scenario, turbulence within turbulence becomes impossible to distinguish observationally.

To address this problem, several analytical studies have sought to provide quantitative descriptions of KHI dynamics.
The series of works began with an idealised configuration of two layers with different densities and oppositely directed velocities \citep{2019ApJ...885..101H}. The resulting velocity shear generates KHI, driving turbulence that mixes plasma and exchanges momentum between layers. Conservation of mass, momentum, and energy in the mixing layer enables analytic predictions of the average density, velocity, and temperature profiles, which have been well-verified by simulation \citep{2019ApJ...885..101H}. Decades of research have established that turbulence layers grow self-similarly \citep{1974JFM....64..775B,PhysRevLett.106.104502}. Starting from these widely accepted results, \citet{2023MNRAS.520.1738H} formulated the dynamics of the mixing layer and benchmarked it using 2.5D MHD simulations. This scenario was later extended to 3D oscillating loop driven by a transverse velocity pulse \citep[][hereafter: \citetalias{2024ApJ...966...68H}]{2024ApJ...966...68H}, revealing two distinct stages of turbulence evolution: a development phase in which the height of the mixing layer grows $\propto t$ until the loop becomes fully turbulent, followed by a decay phase where the turbulence energy decreases $\propto t^2$. 

Based on these results, \citet{2025ApJ...991..208Z} (hereafter \citetalias{2025ApJ...991..208Z}) developed an analytic formula describing the time evolution of the displacement amplitude of impulsively driven transverse loop oscillations.
The model predicts two distinct macroscopic signatures of KHI-induced turbulence damping, namely a time-varying damping rate and a frequency drift or jump as the system approaches the fully turbulent stage. These signatures distinguish nonlinear damping from traditional linear damping models and have been tested against observational data via Bayesian model comparison. 

%Zhong + 2025 developed an analytic formula for the time evolution of nonlinear transverse waves and applied Bayesian model comparison to separate linear and nonlinear regimes.

In this work, we extend the above analytical framework by identifying observational signatures of nonlinear kink oscillations damped by KHI-induced turbulence, using 3D MHD simulations across parameter ranges and synthetic imaging data produced with forward modelling (FoMo; \citealt{2016FrASS...3....4V}). We focus on quantitatively detectable features and the validation of our turbulence damping model. %[MOVE TO CONCLUSION:][In simulations, nonlinear effects include the presence of higher-order modes, period increases, and amplitude reduction, in addition to time-evolving frequency drift and damping rate as revealed in \citetalias{2025ApJ...991..208Z}.These features are preserved in artificial image sequences at the spatial resolution of the Atmospheric Imaging Assembly (AIA; \citealt{2012SoPh..275...17L}), except that higher-order modes require higher spatial resolution for clear identification. We also find that images from different line-of-sights (LoS) and wavelengths preserve the period, but the apparent damping rate varies. Oscillations detected in different channels agree well with the simulated oscillation at the early stage, but they exhibit smaller amplitude and larger phase shift at the later stage when the turbulence is well-developed. Recovering such behaviour in our nonlinear turbulence-damping theory requires an overestimated density parameter and an underestimated boundary parameter.]
%%The remainder of this paper is organised as follows. Section~\ref{sec:method} describes the methods. Results in numerical simulations and forward modelling are presented in Section~\ref{sec:result_simulation} and Section~\ref{sec:result_images}, respectively. Section~\ref{sec:discussion} displays the discussion, and Section~\ref{sec:conclusion} summarises our conclusions.

\section{Methods}\label{sec:method}
\subsection{MHD numerical simulation}\label{sec:method_simulation}
\begin{figure}[ht]
	\centering
	\includegraphics[trim=4cm 3.5cm 4cm 4.5cm,clip=true,width=0.7\linewidth]{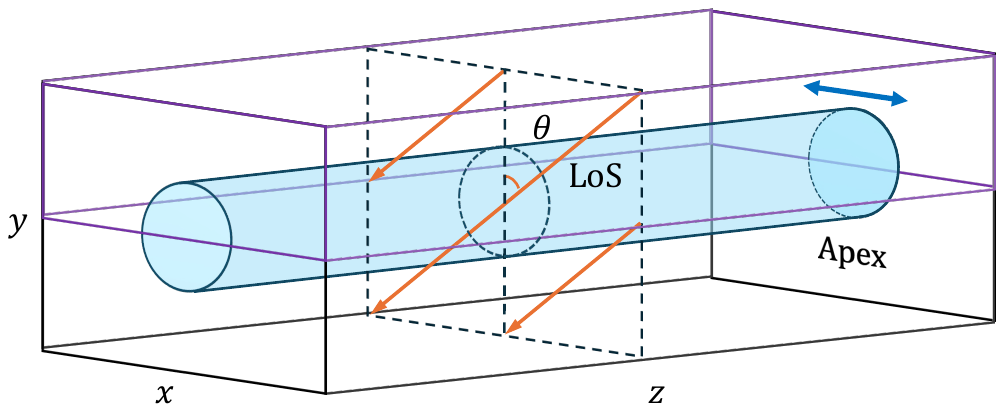}
	\caption{
        The loop in the 3-dimensional (3D) simulation domain $[x,y,z]$ (purple box) and FoMo setting (the whole box, resulting from mirroring the simulation box in the $y$-direction). The loop is set to have a higher density ($\rho_i$, in blue colour) than the surroundings ($\rho_e=1$). The blue arrow indicates the transverse oscillation direction. The dashed rectangle outlines the slice of emissivity across the loop's cross-section. The orange arrows indicate the LoS directions, with their length indicating the integration path to calculate the emission. The $\theta$ denotes the LoS angle with respect to the $y$-axis.
	}
	\label{fig:sketch}
\end{figure}

We numerically simulate an impulsively triggered transverse oscillation of a straight coronal loop with circular cross-section in a 3D domain under an ideal MHD scheme using the P\underline{I}P code \citep{2016A&A...591A.112H}. The governing MHD equations are the same as \citetalias{2024ApJ...966...68H}, and the general numerical setup is similar.
The 3D simulation domain is defined in Cartesian coordinates with $x\in[-L_x,L_x],y\in[0,L_y],z\in[0,L_z] $ where $L_x=2.7,L_y=2.4,L_z=30$, with the length normalised to the loop diameter, and has $[576,512,256]$ grids-points in $[x,y,z]$. Initially, the loop centre is at $x=0,y=0$. To optimise computational resources, we model only half of the loop (from footpoint to apex) with a half cross-section, and then mirror the result in the $y$-direction to obtain the full cross-section.
The modelled loop has a radius $R=0.5$ and internal density $\rho_i$, embedded in a uniform background with lower density $\rho_e=1$ (see Fig.~\ref{fig:sketch}). The transverse density profile is prescribed using a hyperbolic tangent function as in \citetalias{2024ApJ...966...68H}. The loop is surrounded by hotter plasma, such that the internal-to-external temperature ratio satisfies $T_i/T_e<1$.
The internal/external magnetic field $B_i/B_e$ is uniform, respectively, aligned along the $z$-axis. The external sound speed is set to $C_s=1$, plasma $\beta=0.05$, and adiabatic index $\gamma=5/3$. Total pressure is balanced across the loop, so variation in temperature results in corresponding differences in pressure and magnetic field strength ($B_i\ne B_e$). However, the differences in the magnetic field inside and outside the loop are minor. The external and internal Alfv\'en speed are given by $C_{Ae} = \sqrt{\frac{2}{\gamma\beta}}C_s$ and $C_{Ai}=\sqrt{\frac{\rho_e}{\rho_i}}\frac{B_i}{B_e}C_{Ae}$. We consider cases with either $T_i/T_e=0.5$ or $P_i/P_e=3/2$ as control experiments to examine the influence of temperature or pressure imbalance. %[Q:The purpose to set a pressure contrast? allow to generate current?--to stop T being inverse of density, offering larger parameter space.]
An impulsive transverse velocity pulse $v_{x0}$ in the form of 
\begin{equation}
    v_{x0}(x,y,z) = \frac{V_0}{2}\left(1-\tanh{\left(64\left(\frac{\sqrt{x^2+y^2}}{R}-1\right)\right)}\right) \sin\left(\frac{\pi z}{2L_z}\right)
\end{equation}
is applied in the $x$ direction to excite a standing kink oscillation, where $V_0$ is the initial velocity value. To generate large-amplitude oscillations as observed \citep{2019ApJS..241...31N}, we set $V_0$ such that the nonlinearity parameter satisfies $\frac{V_0L}{C_{\rm k}R}\geq1$, where $L=2L_z=60$ is the full loop length and $C_{\rm k}$ is the kink speed from linear MHD wave theory given by
\begin{equation}
    C_{\rm k} = \sqrt{\frac{\rho_iC_{Ai}^2+\rho_eC_{Ae}^2}{\rho_i+\rho_e}}.
\end{equation}

This study investigates the nonlinear effects under varying density contrast $\zeta=\rho_i/\rho_e$. The effect of $T_i/T_e$ is also considered as it influences the telescope response and resulting Extreme Ultraviolet (EUV) emission. Initial conditions for different $\zeta$ cases are listed in Table~\ref{tab:simulation}. %and \textcolor{red}{the theoretically predicted average transverse profiles of density, pressure, and temperature in the mixing layer when its height reaches $0.8R$} are shown in the first column of Fig.~\ref{fig:emission}. 
{And the average transverse profile of density, pressure, and temperature in the mixing layer can be calculated as follows. First, the average density profile is considered to be a third-order polynomial whose coefficients are derived under the assumptions of mass and momentum conservation in the mixing layer \citep{2019ApJ...885..101H}. Given that $C_s =\sqrt{\frac{\gamma P}{\rho}}$ is constant across the loop, the pressure profile is in a similar shape to the density, despite the internal-to-external ratio. Then the temperature profile is obtained according to the ideal gas law. Representative average transverse profiles of these parameters for different values of $\zeta$ when the mixing layer reaches a height of $0.8R$} are shown in the first column of Fig.~\ref{fig:emission}.

The measurement of key parameters in simulation data is stated as follows. The centre-of-mass velocity in the $x$-direction, $V_{\rm CoM}$, is calculated by
\begin{equation}
    V_{\rm CoM}=\frac{\int \rho v_xdxdy}{\int \rho dxdy},
\end{equation} %$V_{\rm CoM}=\frac{\int \rho v_xdxdy}{\int \rho dxdy}$, 
\noindent where $\rho$ and $v_x$ are the density and transverse velocity in each simulation grid respetively, and the loop displacement is a result of the time integral of $V_{\rm CoM}$: 
\begin{equation}
\xi (t) = \int_0^{t} V_{\rm CoM} dt.
\end{equation}
From the time series of $V_{\rm CoM}$, we measure the initial loop velocity {after the kick-off}, $V_i$, as the average velocity over the first trough {to minimise the noise. Here, $V_i$ is smaller than $V_0$ due to the initial impulsive energy leakage into the surroundings, as first described by \citet{2006ApJ...642..533T}.} We also extract the velocity shear $\Delta V$ from the velocity profile across the loop as the velocity difference inside and outside the loop. 
Importantly, we define a density threshold $\rho_T$ to characterise the lower density limit of the effective collective motions. $\rho_T$ close to $\rho_i$ means the loop core motion is dominated, while $\rho_T\rightarrow\rho_e$ indicates the whole mixing layer is included. In other words, the oscillation parameters/evolution are $\rho_T$-dependent. The time variation of mass above a given threshold $\rho_T$ is obtained, from which the mass change rate above a density threshold $ m_{\rho\geq\rho_T}$ is measured (see similar measure in Fig.~10 of \citetalias{2024ApJ...966...68H}). The mixing coefficient $C_1$, a proxy quantifying how fast the KHI-induced mixing goes on, is calculated using Eq.(2) of \citetalias{2025ApJ...991..208Z} with $R,\zeta,\Delta V,m_{\rho\geq\rho_T}$. As an empirical proxy derived from the measured growth rate, $C_1$ ranges from 0.1 to 1 for the current setup. In previous 3D MHD simulations of turbulence mixing, $C_1$ is found to be on the order of 0.3 \citep{2023MNRAS.520.1738H,2024ApJ...966...68H}, compared to approximately 0.5 in  hydrodynamic cases \citep{1974JFM....64..775B}. While $C_1$ is model-dependent, these values provide a benchmark for interpreting the efficiency of momentum exchange across the loop boundary.
%\textcolor{red}{As shown in Table~\ref{tab:simulation}, $C_1$ generally grows with the increasing density contrast, which is unexpected and need further investigation in future study.}
%We measure the mass change rate above a density threshold $ m_{\rho\geq\rho_T}$ and then use it along with $R,\zeta,\Delta V$ to inversely estimate the value of $C_1$ using Eq.(2) of SZ25.
%\Delta m_{\rho>\rho_T} [undefined]

%Define the nonlinearity parameter $\frac{V_iL}{C_{\rm k}R}$.

\begin{table}
    \centering
    \begin{tabular}{cccccccccc}
    \hline
      $\zeta$ & $T_i/T_e$ & $P_i/P_e$ & $B_i/B_e$ & $C_{\rm k}$ & $V_0$ & $V_i$ & $\Delta V$ & $C_1$   \\
      \hline\hline
       2  & 0.5 & 1 & 4.90/4.90 & 4.0 & 0.15 & $0.11\pm0.02$ & $0.11\pm0.02$  & $0.31^{+0.04}_{-0.03}$ & \\
       2  & 0.8 & 3/2 & 4.84/4.90 & 3.97 & 0.15 & $0.11\pm0.02$ & $0.11\pm0.02$  & $0.30^{+0.04}_{-0.04}$ & \\
       3  & 0.5 & 3/2 & 4.89/4.90 & 3.44 & 0.15 & $0.12\pm0.02$ & $0.13\pm0.02$ & $0.39^{+0.08}_{-0.05}$ & \\
       4  & 0.5 & 2 & 4.78/4.90 & 3.06 & 0.10 & $0.08\pm0.01$ & $0.09\pm0.01$ & $0.45^{+0.08}_{-0.06}$ & \\
       5  & 0.5 & 2.5 & 4.72/4.90 & 2.77 & 0.10 & $0.08\pm0.01$ & $0.08\pm0.02$ & $0.49^{+0.11}_{-0.08}$ & \\
       5  & 0.3 & 3/2 & 4.84/4.90 & 2.81 & 0.10 & $0.08\pm0.01$ & $0.08\pm0.02$ & $0.62^{+0.13}_{-0.10}$ & \\
       5  & 0.5 & 2.5 & 4.72/4.90 & 2.77 & 0.15 & $0.12\pm0.03$ & $0.13\pm0.03$ & $0.54^{+0.17}_{-0.10}$ & \\
       5  & 0.3 & 3/2 & 4.84/4.90 & 2.81 & 0.15 & $0.13\pm0.03$ & $0.13\pm0.03$ & $0.58^{+0.16}_{-0.10}$ & \\
       10  & 0.15 & 3/2 & 4.84/4.90 & 2.08 & 0.07 & $0.07\pm0.01$ & $0.07\pm0.01$ & $0.75^{+0.13}_{-0.10}$ & \\
       \hline
    \end{tabular}
    \caption{The initial conditions of the 3D MHD simulations and measured parameters: $V_i$ (initial loop velocity), $\Delta V$ (velocity shear), $C_1$ (mixing coefficient quantifying the rate of KHI-induced mixing).
    }
    \label{tab:simulation}
\end{table}

\subsection{Forward modeling}
\label{sec:method_fomo}
To investigate nonlinear signatures in EUV images at current telescope resolution, we synthesise imaging data using the forward modelling code for optically thin plasma, FoMo \citep{2016FrASS...3....4V}. FoMo calculates the emissivity in each voxel of the 3D simulation domain based on local electron density, temperature, and velocity values. The emissivity is then integrated along a user-defined LoS to produce synthetic intensity images or spectra at a selected channel/wavelength $\lambda$ of a specific telescope. 
In this work, we apply FoMo to render our simulations into images in four EUV channels (131\,\AA, 171\,\AA, 193\,\AA, 211\,\AA) of Atmospheric Imaging Assembly (AIA; \citealt{2012SoPh..275...17L}) on board the Solar Dynamics Observatory (SDO; \citealt{2012SoPh..275....3P}). AIA has provided continuous full-disk solar images since 2010, and is widely used to study coronal loops and wave phenomena. 
We set the loop radius as 1~Mm, external electron density as $10^9$~cm$^{-3}$, and internal temperature $T_i=1$~MK, and all other parameters are scaled accordingly.
We set the LoS perpendicular to the loop's axis (i.e., in the $x-y$ plane; see orange arrows in Fig.~\ref{fig:sketch}), and vary the LoS angle from $0^{\circ}$ (along the $y$-axis, optimal for viewing transverse motions) to $90^{\circ}$ (along the $x$-axis, parallel to the transverse motions). 
The simulation grid has a spatial size of 18.8~km in $x$ and 9.2~km in $y$, significantly finer than AIA's spatial plate scale of 440~km/pixel. We therefore rebin the data to match AIA's effective resolution. 
Then a point spread function (PSF) is applied to the synthetic image to account for AIA's optical response, mainly shaped by instrumental scattering \citep[e.g.,][]{Grigis2011,2013ApJ...765..144P}. Poisson noise is added to mimic observational uncertainty.
Temporal integration is omitted, as the time cadence of the simulation (7.2/9.3/11.8~s in the case of $T_e=1.25/2/3$~MK) is greater than the typical exposure time ($0.5-3$~s) of AIA \citep{2012SoPh..275...17L}.

%which measures the effect of scatter lights,
%Set the view angle by the vector of the line-of-sight and the plane of sky. wavelengths, response function, and resolution.

\begin{figure}
	\centering
	\includegraphics[width=\linewidth]{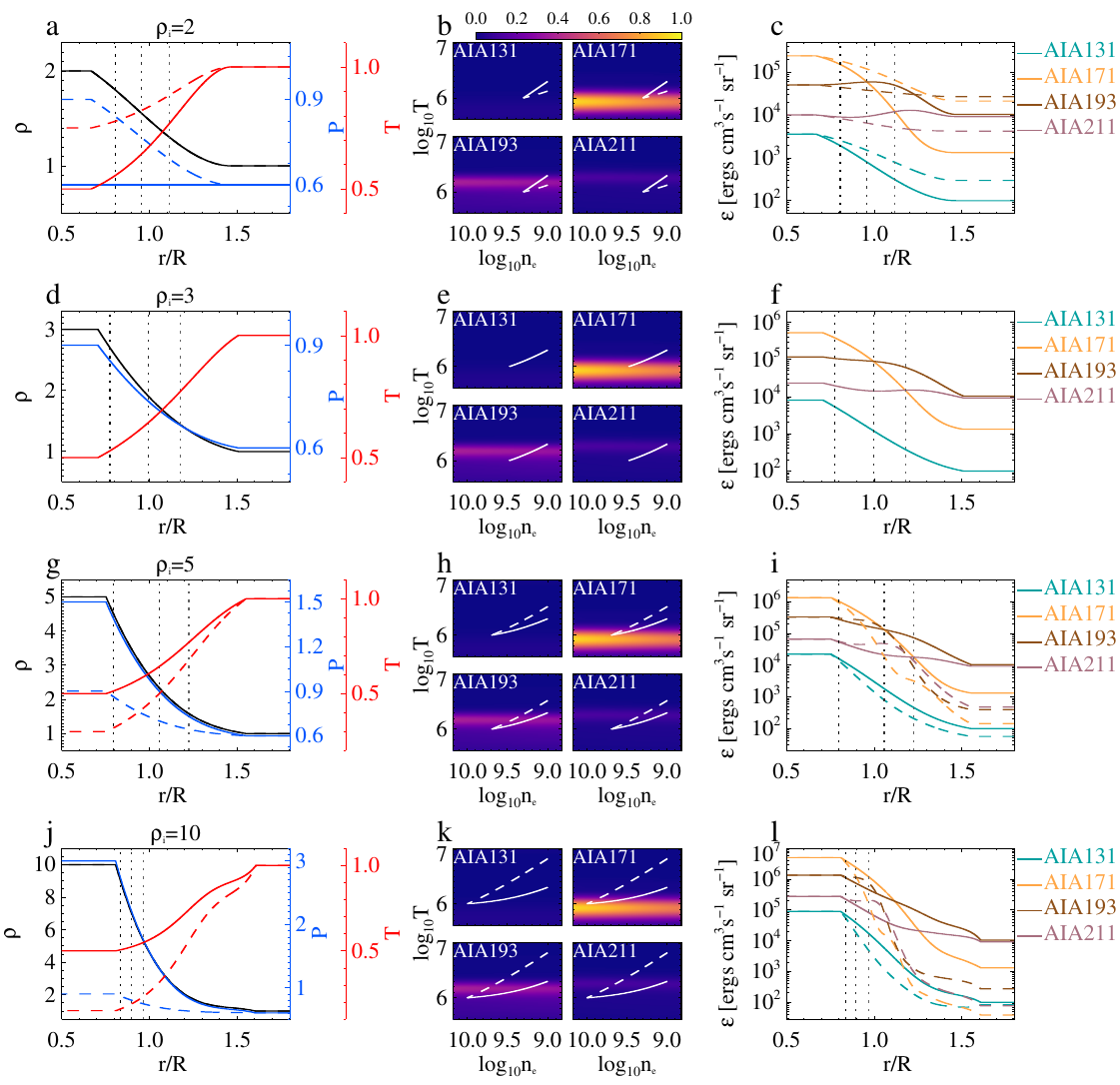}
	\caption{
        Transverse profile of density, pressure, temperature and synthetic emissivity in AIA EUV channels across the loop with different density contrasts when the turbulent layer reaches a height of 0.8$R$. Left: average transverse density (black), pressure (blue) and temperature (red) across the loop. The solid curves represent the case with $T_i/T_e=1/2$ while the dashed curves with $P_i/P_e=3/2$. Middle: contribution function $G(n_{e},T)$ in 131\,\AA, 171\,\AA, 193\,\AA, and 211\,\AA\ of AIA, overlaid with the corresponding loop transverse profile determined by density and temperature on the left column. Right: transverse emissivity across the loop in four channels of AIA. The dotted vertical lines from left to right indicate the location of the upper limit of the turbulent layer, defined by the density threshold $\rho_T=0.9\rho_i$, and where two channels have equal emissivity, respectively. These lines are duplicated in the left panels for reference. 
	}
	\label{fig:emission}
\end{figure}

For cases with varying $\zeta$ and $T_i/T_e$, the corresponding contribution $G_{\lambda}(n_e,T)$ profiles, where $\lambda$ is the observed wavelength and $n_{e}$ is the electron density with $n_e =\rho\times 10^9$cm$^{-3}$, are shown in the second column of Fig.~\ref{fig:emission}, and the resulting emissivity in four AIA channels is demonstrated in the last column. In general, AIA 171\,\AA\ captures the 1-MK-hot loop well, consistent with its peak response temperature of $10^{5.8}$~K. Since the external background is hotter, hotter lines such as 193\,\AA\ and 211\,\AA\ are more sensitive to the loop boundary and ambient medium. As a result, for regions where density approaches $\rho_i$, the emission in 171\,\AA\ dominates over the other channels. Conversely, in regions close to $\rho_e$, hotter channels yield stronger emission. This behaviour agrees with the findings of \citet{2016ApJ...830L..22A}.
It is also worth noting that the density profile does not directly translate into the emissivity profile. As shown in Fig.~\ref{fig:emission}c--i, the density falls off sharply, while the 211\,\AA\ emissivity remains relatively flat. Rather, the 171\,\AA\ emissivity profile resembles the temperature profile, especially the internal-to-external difference by the order of magnitude. This is naturally inherited from the contribution function where the influence of a temperature difference given in our work is bigger than that of the density difference.
This also brings the question about whether the motion of the centre of mass (CoM) can represent that of the centre of emission (CoE), which is discussed in Section~\ref{sec:CoM_vs_CoE}.

%the weight of density is not linearly inherited in the emissivity profile, see Fig.~\ref{fig:emission}c--i, the density profile is decreasing sharply while the emissivity in 211\,\AA\ is flat. 

%Fig.~\ref{fig:emission} shows 171 captures the loop cores' dynamics, while 193 and 211 are sensitive to the loop boundary/external medium.

\section{Results}

For impulsively driven transverse oscillations of a thin loop ($R\ll L$) with nonlinearity parameter $\frac{V_0L}{C_{\rm k}R}>1$, nonlinear effects become significant. This section aims to describe nonlinear signatures found in the behaviour of large-amplitude decaying transverse loop oscillations damped by KHI-induced turbulence revealed in high-resolution 3D MHD simulations (Section~\ref{sec:result_simulation}) and in observational signatures detected in the corresponding synthetic EUV imaging data (Section~\ref{sec:result_images}) produced using the FoMo forward modelling code. 
In both simulations and theoretical analyses, the loop's motion is tracked via its centre of mass, whereas in imaging observations, the motion is inferred from the Gaussian centre of emission. The consistency and discrepancies between these two approaches are discussed in Section~\ref{sec:CoM_vs_CoE}.
%[In this study, the loop boundary is assumed to be sharp and thin to suppress resonant absorption, allowing us to focus primarily on the nonlinear damping induced by KHI.][MOVE BELOW]

\subsection{Nonlinear signatures in simulations}\label{sec:result_simulation}
%In this section, we present the nonlinear signatures of large-amplitude transverse loop oscillations in a coronal loop, as obtained from high-resolution 3D MHD simulations. These oscillations are damped via KHI-induced turbulence. [Several damping mechanism works, among which KHI-induced turbulence dominates in some conditions.]

%A brief review of previous results [move to Introduction?]
%First is a brief review of prior works on this topic. The studies of KHI dynamics started with an idealised configuration: two layers with different densities and opposite velocities \citep{2019ApJ...885..101H}. The velocity shear between the layers generates KHI and forms turbulence, leading to density mixing and momentum exchange. The conservation of mass, momentum, and energy in the mixing layer allows analytic predictions for average density, velocity, and temperature profiles, which is well verified by simulation \citep{2019ApJ...885..101H}. And the turbulence layer was shown to grow self-similarly \citep{2023MNRAS.520.1738H}. This turbulence developing scenario was extended to 3D oscillating loop driven by a transverse velocity pulse \citep{2024ApJ...966...68H}. That work identified two distinct stages in turbulence evolution: a development phase with layer height growth $\propto t$ until the loop becomes fully turbulent, followed by a decay phase where the turbulence energy decreases $\propto t^2$. 

Numerical simulations performed by \citetalias{2024ApJ...966...68H} showed that nonlinear kink waves can host KHI, triggered by the velocity shear at the loop surface, which subsequently creates turbulence in the loop. The adopted sharp and thin loop boundary prevents resonant absorption, and the KHI forms a mixing layer around the loop core. The self-similar nature of the growth of the mixing layer allowed \citetalias{2025ApJ...991..208Z} to derive an analytic formula for the time evolution of a transversely oscillating loop damped by KHI-induced turbulence. The formula provides the mass and momentum of the oscillating loop (divided into core and mixing layer) and hence the oscillation amplitude, under certain assumptions. 
This analytic model incorporates parameters such as loop radius $R$, density contrast $\zeta$, initial velocity $V_0$, velocity shear $\Delta V$, kink period $\rm P_{k}$, mixing coefficient $C_1$ {and density threshold $\rho_T$} (see Eq.12 in \citetalias{2025ApJ...991..208Z}). {The value of $\rho_T$, ranging from the external and internal density, defines the lower limit of the region participating in the collective oscillation. A smaller $\rho_T$ (closer to the external density) implies that a larger fraction of the mixing layer is included, leading to stronger damping of the oscillation.}

In this model, the loop core is an impulsively driven oscillator at the natural kink frequency, its motions damp at the same rate as the mixing layer grows. Meanwhile, the mixing layer acts as a forced oscillator driven by the motion of the loop core, oscillating at the average of kink and characteristic Alfv\'en frequency of the mixing layer, modulated by an envelope at half of the difference between the kink and Alfv\'en frequency.
As a result, the collective motion of the loop is the interference of two cosine waves, one at the kink frequency and the other at the oscillation frequency of the mixing layer. As turbulence develops, the oscillation frequency gradually drifts from the kink frequency towards the mixing layer frequency. In addition, at the early stage, the damping is weak because the size of turbulence is small, and as it grows, the damping strengthens. This results in a time-evolving damping rate.

Our simulations explore nonlinear effects of large-amplitude transverse oscillations of a straight coronal loop under varying $\zeta, T_i/T_e$, and initial driving velocity $V_0$. The setup satisfies the long wavelength condition, with $R/L=0.5/60<0.01$. 
The perturbation set in this work ($\frac{V_0L}{C_{\rm k}R}\in(0,9]$) is stronger than the in previous simulations \citep[e.g., $\frac{V_0L}{C_{\rm k}R}\leq4,=1.9,=2.6$ in][respectively]{2016A&A...595A..81M,2017ApJ...836..219A,2019FrP.....7...85A}. In real observations, kink oscillation velocity amplitudes can reach up to 300~km~s$^{-1}$ \citep{2019ApJS..241...31N}, corresponding to a nonlinearity parameter exceeding 15. In our simulations, the adopted parameters fall within this observationally relevant nonlinear regime. %say, $\xi_0=0.5R$ in \citet{2016A&A...595A..81M}, $A_0L/R=3.3$ in \citet{2019FrP.....7...85A}. v0L/C_{\rm k}R=1.9 in Antolin+2017.

In simulations, once the loop is impulsively driven, velocity shear at the boundary triggers KHI, forming a turbulent layer that grows linearly with time. This growth depends on the root-mean-square velocity and $C_1$, which agrees well with the theoretical model (see Appendix~\ref{append:1} and Supplementary animation therein).
Example average profiles of density, pressure, and temperature are shown in Fig.\ref{fig:emission}. Additionally, we observe compression of the loop cross-section at the oscillation crests/troughs (see animation), which results from a combination of higher-order modes $m\geq2$ and inertia of dense loop itself \citep{2014SoPh..289.1999R,2019FrP.....7...85A}. %\textcolor{red}{[Q: How to distinguish the compression is caused by $m>2$ modes or by the inertia? Impossible to tangle. It is the inertia that generates m>2]}
%In the nonlinear regime, the oscillation mode is no longer a pure kink mode but a combination of multiple modes \citep{2014SoPh..289.1999R,2019FrP.....7...85A}. 

The turbulence extracts momentum from the loop core to the boundary, damping the oscillations. We track the displacement of the loop's centre of mass (CoM), normalised by 1 in length and by respective linear kink period ${\rm P_k}$ in time, to enable comparison across cases. As shown in Fig.~\ref{fig:simul_curve}a--b, oscillations are grouped by temperature contrast $T_i/T_e=1/2$ and pressure contrast $P_i/P_e=3/2$. The oscillations are non-sinusoidal and damp faster with higher nonlinearity, quantified by $\frac{V_iL}{C_{\rm k}R}$. Consistent with \citetalias{2025ApJ...991..208Z}, we observe frequency drift and a time-evolving damping envelope. 
%Examining the period variation and decay pattern, damping oscillations of the loop by turbulence exhibit common characteristics, including frequency drift towards longer periods and a time-evolving damping rate, as revealed in SZ25. 

\subsubsection{Period increase and amplitude decrease}
\begin{figure}[ht]
	\centering
	\includegraphics[width=0.95\linewidth]{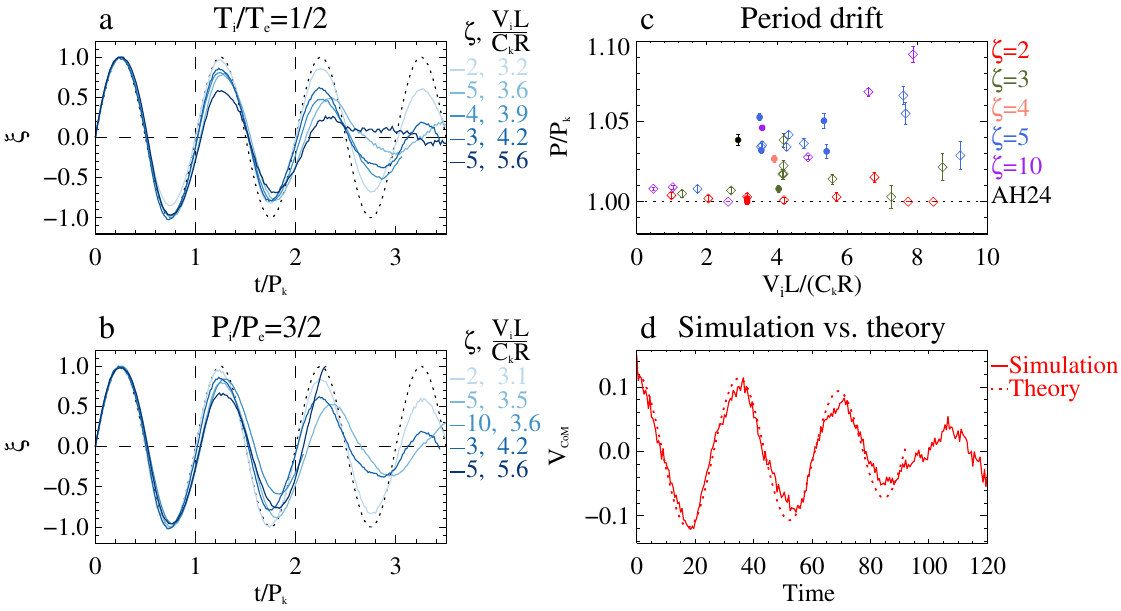}
	\caption{
        Amplitude-dependent oscillation period deviation from the linear kink period $\rm P_{\rm k}$ (a--c) and amplitude discrepancy (d) measured in simulation data. The dotted curves are a standard stationary sine wave for reference. (a--b): Time series of displacement amplitude $\xi_{\rho\geq0.9\rho_i}$ measured in two sets of simulations: $T_i/T_e=1/2$ (a) and $P_i/P_e=3/2$ (b). (c): The ratio of measured base period ($\rm P$) and linear kink period ($\rm P_{\rm k}$) as a function of nonlinearity parameter. Different colours indicate different density contrasts ($\zeta$). The filled symbols indicate data measured from the high-resolution simulations, otherwise from low-resolution simulations taken from \citetalias{2025ApJ...991..208Z}. (d): The comparison of centre-of-mass velocity $V_{\rm CoM,~{\rho\geq0.9\rho_i}}$ measured in simulation (solid) and in theory (dotted) for the case of $\rho_i=3$ shows the former has a lower amplitude than expected. The theoretical curves are computed using our turbulence-damping theory (Eq.12 of \citetalias{2025ApJ...991..208Z}) with input parameter values measured in the simulations. %\textcolor{red}{[a--b legends: REPLACE with $V_0L/C_kR$?]}
	}
	\label{fig:simul_curve}
\end{figure}

Of note, the oscillation period exceeds ${\rm P_k}$ by a few percent, see the curves at around $t/ \rm{P_k}=1$, with the discrepancy increasing with amplitude. 
%[To quantify this effect, we determine the base period ${\rm P}$ ]
To quantify this period discrepancy, for each simulation's time series, we calculate the difference between the period expected from linear theory ${\rm P_k}$, and the fitted base period ${\rm P}$ via curve fitting with Eq.12 of \citetalias{2025ApJ...991..208Z}, using \texttt{mpfitfun.pro} and fixed parameters ($R, L, \zeta, C_1, B_i/B_e$) measured in simulations.
We do not consider time-resolved frequency analysis, as it is affected by the time-evolving frequency drift and limited by the short duration (often $<3$ cycles), resulting in uncertainty greater than 10\% of the base period. To increase the samples for statistics, we also include low-resolution simulations from \citetalias{2025ApJ...991..208Z}.
As shown in Fig.~\ref{fig:simul_curve}c, the amplitude-dependent period increase is negligible (1\%) for $\frac{V_iL}{C_{\rm k}R}\leq2$, but can reach 10\% for larger amplitudes. However, for extreme amplitudes ($\frac{V_iL}{C_{\rm k}R}>7$), the fit fails, likely due to dominance of higher-order modes or nonlinear coupling beyond the turbulence-induced damping model.
The physical origin of this period increase remains unknown and is beyond the scope of this work, warranting further investigation to clarify its underlying cause. %Here, we aim to empirically characterise the nonlinear signatures rather than develop a full theoretical explanation ({\bf\em no need to provide excessive justification}). 
Nonetheless, this effect highlights the importance of measuring the actual period in simulations or observations, rather than assuming the linear kink-mode period.

In addition to period discrepancy, the oscillation amplitude is found to be lower than the theoretical predictions of the analytical model in \citetalias{2025ApJ...991..208Z}. As displayed in Fig.~\ref{fig:simul_curve}d, the CoM velocity, $V_{\rm CoM}$, measured above a density threshold $\rho_T$, has an amplitude around 10\% smaller than theoretically expected (see comparison between the solid and dotted curve at the maxima and minima). The displacement (time integral of $V_{\rm CoM}$) consequently is lower. This may be due to part of the energy being transferred to higher-order, smaller-scale modes, especially those responsible for the observed cross-section deformation. %\textcolor{red}{Furthermore, in a bent loop, the existence of $m\geq 2$ modes breaks the symmetry of ?}.
An additional difference between the simulations and our analytical model is that, in the later stage when the kinetic energy of the loop core approaches that of the mixing layer, the turbulence develops more quickly and consequently the simulated oscillation decays faster than the theoretical expectation calculated using the measured parameters in simulations. As a result, the loop arrives at the fully turbulent stage earlier. In fact, this behaviour is consistent with the prediction already discussed in Section 3.2 of \citetalias{2024ApJ...966...68H}, where it was shown that once the assumption of core energy dominance no longer holds, the accuracy of the model naturally decreases. Thus, the observed difference is not a shortcoming of the model but rather a regime where its underlying assumptions no longer apply, as anticipated. An example is shown in the supplementary animation for the case of $\zeta=3$, where the turbulent layer penetrates deeper into the core than the model since $t=40$ and completed at $t\approx80$, compared to the theoretically estimated $t=104$. This faster damping in the later stage is more evident for cases with greater nonlinearity.

%Inputting the measured values of parameters into the turbulence-damping theory always gives broad consistency with the simulated oscillation, with minor discrepancies ({\bf\em we say minor here and notable in the previous paragraph}) mentioned above. To make the model best fit the simulated oscillations, values in a combination of $C1/R,\zeta,\rho_T$ derivate while others stay the same as the measured, as revealed in the Appendix~\ref{append:A} and Table~\ref{tab:fit_CoM} therein. Specifically, an underestimated $C1/R$, an overestimated $\zeta$, and a smaller $\rho_T$ jointly produce the reduced amplitude at each crest and the observed faster decay at the later stage. Alternative combination of a greater $\zeta$ and a smaller $\rho_T$ also works in a limited cases. {\bf\em I think this paragraph needs to be improved}

\subsubsection{Model fitting with simulated oscillations}\label{sec:model_fit}
\begin{figure}[ht]
	\centering
    \includegraphics[width=\linewidth]{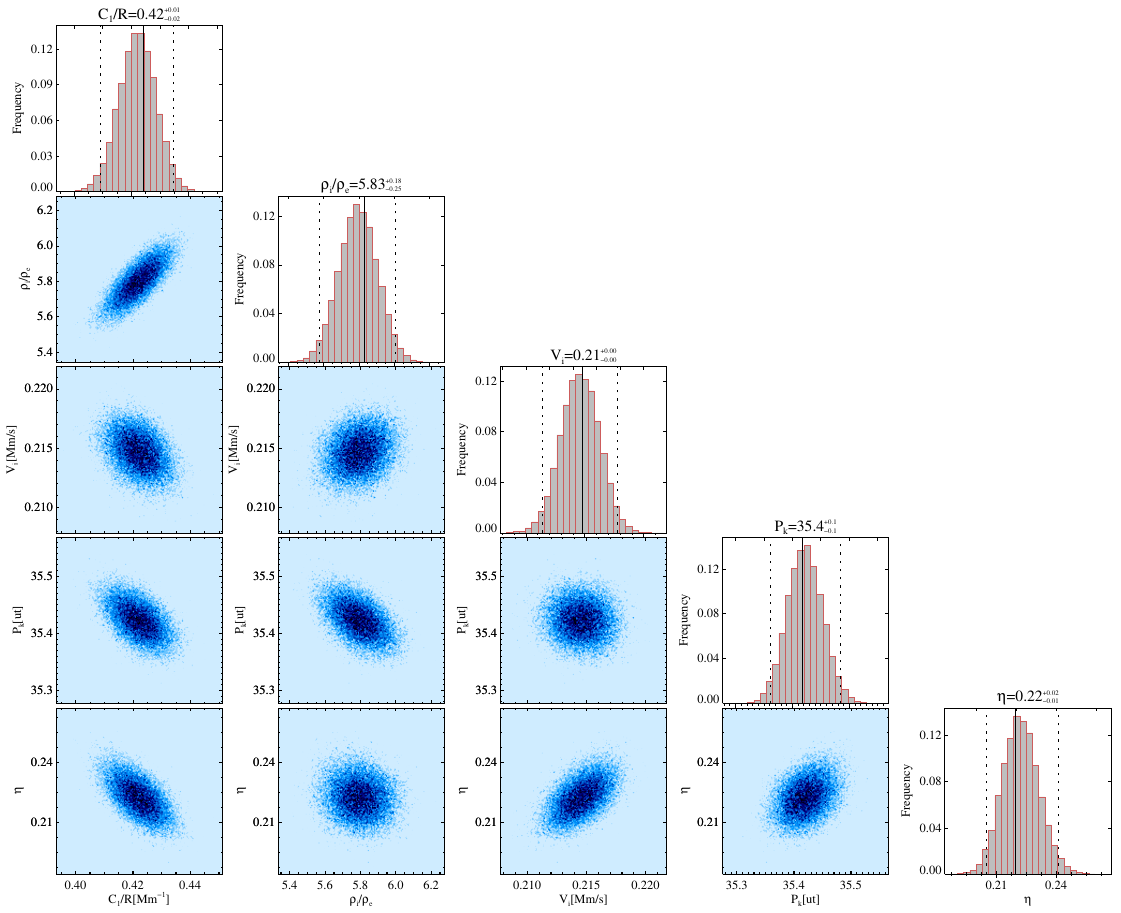}
	\caption{ 
        Corner plot showing the posterior distributions of the parameters in the nonlinear model (Eq.\ref{eq:com_fit}). The diagonal panels display the posterior distributions for individual parameters, with the black vertical lines marking the best-fitted values. The panels below the diagonal show the joint posterior distributions, highlighting correlations between parameter pairs. The data are obtained via MCMC sampling for the case $\zeta=3,T_i/T_e=0.5,V_0=0.15$ with full time series. Elongated contours indicate strong correlations, while circular or rectangular contours suggest weak or no correlation. The unit of ${\rm P_k}$ (\lq\lq ut\rq\rq) is the unit time of simulation.
    }
	\label{fig:corner}
\end{figure}

To quantitatively test if the turbulence-damping function agrees with the simulated oscillations, we apply Solar Bayesian Analysis Toolkit (SoBAT; \citealt{2021ApJS..252...11A}) to fit the oscillations with the function given by
\begin{equation}\label{eq:com_fit}
    \xi_{\rm{CoM}_{\rho\geq\rho_T}} = F(C_1/R,\zeta,V_i, \mathrm{P_k},\eta).
\end{equation}
\noindent where $F$ is the function defined in Eq.(12) of \citetalias{2025ApJ...991..208Z} and $\eta$ is related to $\rho_T$,  representing the fraction of total mass per unit width between the core and the mixing layer (\citetalias{2025ApJ...991..208Z}). 
The fitting results are displayed in Table~\ref{tab:fit_CoM}, where we translate $\eta$ into corresponding $\rho_T$. In the function, $V_i$ mainly determines the initial displacement amplitude, and $\rm{P_k}$ sets the oscillation periodicity. Thus, these two parameters are robustly well-constrained by the oscillatory pattern. Indeed, the values of $V_i$ and $\rm{P_k}$ agree with the measured values and exhibit very narrow distributions (see the seventh--eighth columns in Table~\ref{tab:fit_CoM}). The remaining parameters are correlated and jointly shape the decay pattern. In particular, $C_1/R$ is a mixing term that stands for the efficiency of the density mixing process, $\zeta$ represents loop mass, and $\rho_T$ denotes the upper density limit of the mixing layer. Therefore, a greater $C_1/R$, a smaller $\zeta$ and a smaller $\rho_T$ (equivalently greater $\eta$) individually lead to faster damping, and their reverse causes slower decay. It should be noted that a small increment of $\zeta$ within 1 makes a minor phase difference in the last oscillation cycle, maybe with a slightly smaller displacement. Notably, the three effects tend to cancel each other out. Their coupling is complicated, but in general, we obtained that $C_1/R$ and $\zeta$ are positively correlated while $C_1/R$ and $\eta$ are negatively correlated (see Fig.~\ref{fig:corner} as a representative).

\begin{table}
    %\centering
    \begin{tabular}{ccccccccc}
    \hline
    \multicolumn{2}{c}{Simulation parameters} & & \multicolumn{5}{c}{Fitted parameters} \\ \cline{0-1} \cline{5-9}
     $V_0(\frac{V_0L}{C_{\rm k}R})$ & $T_e/T_i$ & & Segment & $C_1/R$ & $\zeta$ & $V_i$[km~s$^{-1}$] & $\rm{P_k}$~[s] & $\rho_{T}/\rho_i$  \\
      \hline\hline
    0.15 (4.50) & 2 & Sim. values & & $0.31^{+0.04}_{-0.03}$ & 2 & $24\pm4$ & 279 & 0.9 \\
     &    & Fitted values & Full & $0.44^{+0.01}_{-0.02}$ & $6.0^{+0.4}_{-0.3}$ & $21^{+1}_{-1}$ & $280^{+0}_{-0}$  & $0.87^{+0.06}_{-0.07}$  \\ 
     &  & {prior $\zeta\in[1,5]$} & Full & $0.29^{+0.01}_{-0.01}$ & $1.3^{+0.1}_{-0.1}$ & $21^{+1}_{-1}$ & $279^{+0}_{-0}$  & $0.97^{+0.11}_{-0.07}$  \\ \hline
     0.15 (4.53) & 1.25 & Sim. values & & $0.30^{+0.04}_{-0.04}$ & 2 & $19\pm3$ & 353 & 0.9 \\
      & & Fitted values & Full & $0.45^{+0.01}_{-0.02}$ & $5.8^{+0.3}_{-0.2}$ & $17^{+1}_{-1}$ & $356^{+1}_{-1}$  & $0.90^{+0.05}_{-0.06}$ \\ \hline
    0.15 (5.23) & 2 & Sim. values & & $0.39^{+0.08}_{-0.05}$ & 3 & $26\pm4$ & 327 & 0.9 \\
     & & Fitted values & Full & $0.42^{+0.01}_{-0.05}$ & $5.8^{+0.2}_{-0.3}$ & $23^{+0}_{-0}$ & $329^{+1}_{-1}$ & $0.80^{+0.05}_{-0.04}$  \\
     &      &   & T1 & $0.42^{+0.02}_{-0.02}$ & $5.3^{+0.3}_{-0.2}$ & $23^{+0}_{-0}$ & $330^{+1}_{-1}$ & $0.81^{+0.06}_{-0.08}$  \\
     &      &   & T2 & $0.44^{+0.06}_{-0.10}$ & $5.2^{+0.4}_{-0.5}$ & $23^{+0}_{-0}$ & $329^{+1}_{-2}$ & $0.84^{+0.16}_{-0.30}$ \\ \hline
    0.10 (3.92) & 2 & Sim. values &  & $0.45^{+0.08}_{-0.06}$ & 4 & $17\pm2$ & 372 & 0.9 \\
    & & Fitted values & Full &$0.42^{+0.06}_{-0.12}$ & $5.8^{+0.4}_{-0.5}$ & $17^{+1}_{-1}$ & $374^{+1}_{-1}$ & $0.65^{+0.16}_{-0.31}$ \\
    &  & & T2 &$0.41^{+0.07}_{-0.01}$ & $1.0^{+0.5}_{-0.0}$ & $17^{+1}_{-1}$ & $370^{+1}_{-1}$ & $0.99^{+0.01}_{-0.33}$ \\ \hline
    0.10 (4.27) & 2 & Sim. values &  & $0.49^{+0.11}_{-0.08}$ & 5 & $17\pm2$ & 422 & 0.9 \\
    & & Fitted values & Full &$0.42^{+0.01}_{-0.01}$ & $5.7^{+0.2}_{-0.2}$ & $18^{+0}_{-0}$ & $423^{+1}_{-1}$ & $0.64^{+0.05}_{-0.05}$ \\
     &  &  & T2 &$0.61^{+0.14}_{-0.20}$ & $8.5^{+1.3}_{-1.4}$ & $17^{+0}_{-0}$ & $416^{+2}_{-1}$ & $0.84^{+0.24}_{-0.25}$ \\ \hline
    0.10 (4.33) & 3.3 & Sim. values & & $0.62^{+0.13}_{-0.10}$ & 5 & $22\pm3$ & 323 & 0.9 \\
    & & Fitted values & Full &$0.38^{+0.02}_{-0.02}$ & $5.2^{+0.2}_{-0.2}$ & $22^{+1}_{-1}$ & $326^{+1}_{-1}$ & $0.69^{+0.06}_{-0.07}$ \\
    &  & & T1 &$0.85^{+0.06}_{-0.06}$ & $8.6^{+1.1}_{-1.0}$ & $22^{+1}_{-1}$ & $318^{+1}_{-1}$ & $0.96^{+0.04}_{-0.13}$ \\
    &  &   & T2 &$0.96^{+0.04}_{-0.09}$ & $9.2^{+0.7}_{-1.5}$ & $22^{+1}_{-1}$ & $316^{+1}_{-1}$ & $0.90^{+0.12}_{-0.23}$ \\ \hline
    0.15 (6.50) & 2 & Sim. values &  & $0.54^{+0.17}_{-0.10}$ & 5 & $26\pm6$ & 422 & 0.9 \\
    & & Fitted values  & Full &$0.37^{+0.02}_{-0.02}$ & $6.1^{+0.3}_{-0.4}$ & $25^{+1}_{-1}$ & $425^{+3}_{-3}$ & $0.56^{+0.09}_{-0.09}$  \\
    &   &  & T1 & $0.57^{+0.07}_{-0.19}$ & $8.6^{+0.3}_{-0.4}$ & $24^{+1}_{-1}$ & $415^{+5}_{-2}$ & $0.86^{+0.22}_{-0.31}$ \\
    &    &   & T2 & $0.36^{+0.14}_{-0.17}$ & $8.3^{+1.5}_{-1.5}$ & $24^{+1}_{-1}$ & $417^{+3}_{-3}$ & $0.60^{+0.35}_{-0.49}$  \\ \hline
     0.15 (6.40) & 3.3 & Sim. values & & $0.58^{+0.16}_{-0.10}$ & 5 & $36\pm8$ & 323 & 0.9 \\
      %\multicolumn{2}{c}{Fitting results}   & Full &$0.30^{+0.01}_{-0.01}$ & $4.9^{+0.1}_{-0.1}$ & $0.1^{+0}_{-1}$ & $45^{+2}_{-2}$ & $0.19^{+0.08}_{-0.08}$  \\
     & & Fitted values & T1 & $0.15^{+0.14}_{-0.03}$ & $9.4^{+0.6}_{-1.3}$ & $32^{+9}_{-2}$ & $318^{+1}_{-1}$ & $0.12^{+0.63}_{-0.02}$ \\
     &     &   & T2 & $0.30^{+0.12}_{-0.14}$ & $8.9^{+1.0}_{-1.5}$ & $32^{+1}_{-1}$ & $319^{+1}_{-1}$ & $0.58^{+0.34}_{-0.47}$ \\ \hline
     0.07 (4.04) & 6.7 & Sim. values &  & $0.75^{+0.13}_{-0.10}$ & 10 & $27\pm4$ & 309 & 0.9 \\
     &  & Fitted values & Full & $0.97^{+0.03}_{-0.03}$ & $12.1^{+0.6}_{-0.5}$ & $25^{+2}_{-2}$ & $309^{+1}_{-1}$ & $0.87^{+0.05}_{-0.05}$ \\
       \hline
    \end{tabular}
    \caption{The best fitted parameters for CoM oscillations in simulations fitted with the turbulence damping function (Eq~\ref{eq:com_fit}). Note here the parameters' units are converted to those in the forward modelling analysis for comparison to Table~\ref{tab:fit}. Uncertainties are reported as 95\% credible intervals of the posterior distribution. For each case, the first row lists the setup conditions ($T_e, V_0 (\frac{V_0L}{C_{\rm k}R}),\zeta$) and measured values ($C_1/R,V_i,\rm{P_k}$) in the corresponding simulation, while the remaining rows display the fitted values. The column of \lq\lq Seg\rq\rq, denotes the selected time segment of the input oscillation, with \lq\lq Full\rq\rq\ indicate the full time series, \lq\lq T1\rq\rq\ for truncating the oscillations at when the core mass stops linear loss and \lq\lq T2\rq\rq\ for truncating the oscillations at when the kinetic energy of the loop core equals that of the mixing layer.}
    \label{tab:fit_CoM}
\end{table}

Looking at Table~\ref{tab:fit_CoM}, most of the best fits give higher $\zeta$, sometimes with a smaller $\rho_T$ and/or biased $C_1/R$, while other parameters are consistent with the measurement in simulations. The following combinations also appear in one perspective case: a smaller $\zeta$ and a higher $\rho_T$, a greater $C_1/R$ and a smaller $\zeta$. All these possible combinations are caused by the degeneracy between the parameters, as revealed by Fig.~\ref{fig:corner}. We suggest setting a narrower prior distribution of any one of the three parameters (if known) to avoid unrealistic posteriors. See an example in the fifth row, setting the prior of $\zeta\in[1,5]$ gives more reasonable outcomes than setting $\zeta\in[1,10]$ (the second row). %{Anyway, the misestimated parameters indicate that the model does not perfectly agree with the simulations.}
%Specifically, an underestimated $C_1/R$, an overestimated $\zeta$ and a smaller $\rho_T$ jointly produce the reduced amplitude at each crest and the observed faster decay at later stage. Alternative combination of a greater $\zeta$ and a smaller $\rho_T$ also works in a limited cases.
Anyway, the turbulence-damping model reproduces the simulated oscillations with broad consistency, despite minor deviations in regimes where the underlying assumptions are less valid. To achieve the best agreement with the simulations, several model parameters must be adjusted from their measured values in a coordinated manner due to the model degeneracy described above. In particular, a slightly underestimated $C_1/R$, an overestimated $\zeta$, and a smaller $\rho_T$ together yield the reduced amplitude at each oscillation crest and the faster decay observed at the later stage. An alternative combination of a greater $\zeta$ and a smaller $\rho_T$ also reproduces the results in a limited number of cases.

%summary
To summarise, our MHD simulations reveal several key signatures of nonlinear, large-amplitude, decaying transverse loop oscillations. The existence of high-order modes, evident in the dynamic squashing of the loop cross-section, leads to reduced oscillation amplitude compared to theoretical expectations. More interestingly, the oscillation period exceeds the linear kink period in the long-wavelength limit by a few percent, and this increase is amplitude-dependent.
The development of KHI and consequent turbulence results in time-varying frequency drift and damping rate, producing a distinctly non-sinusoidal oscillatory pattern. These effects must be taken into account when interpreting large-amplitude loop oscillations in both simulations and observations. 

%{\bf\em this summary is OK. Only one thing is not clear to me, some signatures were already present in the previous study and some new ones are due to the stronger nonlinearity of the current simulations? [sz: period increase has presented in AH24 but it is too minor to be noticed, the faster decay than modelling in the later stage is expected and explained in AH24. Others are due to stronger nonlinearity.] Another question I have: is the strong nonlinear cases, is still the growth of the mixing layer linear in time? [SZ: yes at the earlier stage where the loop deformation is not strong] Does the nonlinear coupling to azimuthal harmonics somehow interfere with that progressive growth?[SZ: Good idea. I think so but not sure how to quantify this. IA: thank you.]}

\begin{figure}
	\centering
	\includegraphics[width=\linewidth]{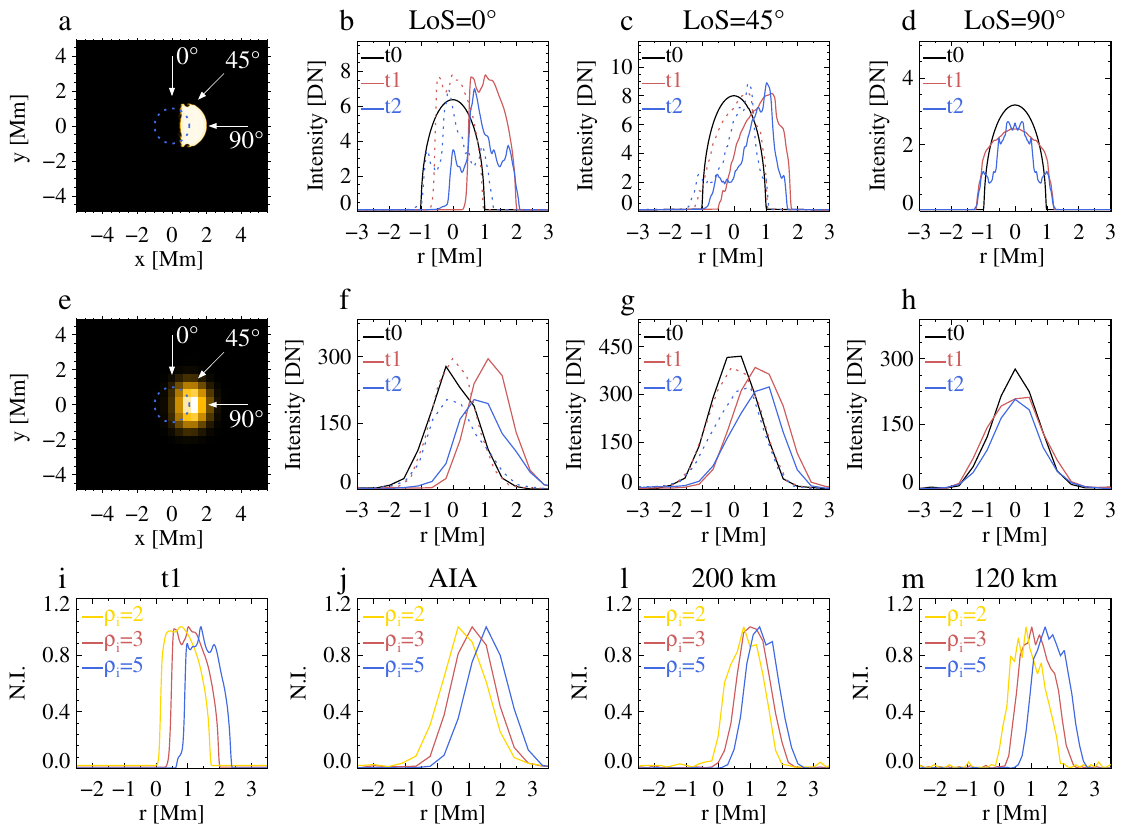} %trim=0cm 4.7cm 0cm 0cm,clip=true,
	\caption{
         Synthetic image of the cross-section of the loop apex in 171~\AA\ channel and transverse intensity profile in different LoS at different spatial resolutions to show the existence of high-order modes ($m\geq2$) generated by nonlinearity. First row (a--d): synthetic data in original simulation resolution. The dotted circle indicate the initial loop's cross-section. The profiles are extracted at three time frames: initial (t0), the first oscillation crest (t1) and the second crest (t2). The dotted curves are the intensity profile in corresponding colours shifted back to the initial location for comparison. Second row (e--h): similar to (a--d), but are degraded into AIA resolution. Deformation of the loop cross-section, manifested as the loop width variation, is invisible in low resolution. Third rows (i--m): The effect of velocity perturbation at t1 in different resolutions, from original, AIA (440~km/pixel), 200~km/pixel to 120~km/pixel. The transverse intensity is normalised (denoted by ``N.I.") for comparison.  %$V_i/C_{\rm k}=[0.037, 0.043, 0.05]$ for $\rho_i=2,3,5$. 
	}
	\label{fig:profile_evo}
\end{figure}

\subsection{Nonlinear signatures in EUV Images} \label{sec:result_images}
We study the EUV imaging signatures of nonlinear decaying kink oscillations of coronal loops damped by KHI-induced turbulence. To achieve this, we apply FoMo to simulation data using the methods described in Section~\ref{sec:method_fomo} to synthesise EUV images at various spatial resolutions and LoS angles, across multiple channels. Here, the LoS angle varies from $0^{\circ}$ to $90^{\circ}$, and the spatial resolution ranges from the original simulation resolution to that of AIA (440~km/pixel), for assessing how resolution affects the visibility of nonlinear features. We focus on the wavelengths that are sensitive to coronal plasma at around 1~MK, including the 131\,\AA, 171\,\AA, 193\,\AA, 211\,\AA\ channels.

\subsubsection{Existence of higher-order modes}
The presence of higher-order modes is also evident in EUV imaging data. The compression of the loop manifests as a squashed cross-section (see Fig.~\ref{fig:profile_evo}a and e), resulting in a time-varying loop width when viewed from a specific LoS. In particular, compression leads to a narrower loop width and enhanced intensity when the LoS is normal to the oscillation direction.
As shown in Fig~\ref{fig:profile_evo}b--d, the width of the transverse intensity profile at the first oscillation crest (red) is narrower than the initial profile (black) for the $0^\circ$ LoS (panel b). The profile also becomes asymmetric, with the peak intensity exceeding the initial value. In contrast, for the $90^\circ$ LoS (panel d), the loop width increases slightly and the intensity decreases. However, the variation in loop width is invisible in degraded resolution. Fortunately, the asymmetry of the profile is detectable in the 0$^{\circ}$ and 45$^{\circ}$ LoS angles with current instrumentation.
%{\bf\em so the changes in the width are not detectable with current instrumentation. Is the same true with the asymmetry of the profile?--[sz: can be reserved in some LoS angles] IA: thank you.}
The compression becomes more pronounced as the oscillation amplitude increases. As illustrated in Fig~\ref{fig:profile_evo}i, the Full Width at Half Maximum (FWHM), obtained by Gaussian fitting at original resolution, decreases with higher amplitudes: for $V_iL/(C_{\rm k}R)=3.2,4.2,5.6$, the FWHM values are 1166~km, 1132~km, and 1126~km, respectively. At AIA resolution, the corresponding measured FWHM increases to 1582~km, 1509~km and 1489~km. At a spatial resolution of 120~km/pixel, the measured FWHM is 1252, 1232, and 1209~km, respectively.

In addition, the KHI vortices appear as thread-like structures in the synthetic EUV images, and higher-amplitude oscillations generate more prominent small-scale features (Fig~\ref{fig:profile_evo}i). Unfortunately, these structures are invisible at AIA resolution (see Fig~\ref{fig:profile_evo}j and o). By gradually improving the spatial resolution (Fig~\ref{fig:profile_evo}j--m, o--q), we find that a minimum resolution of 120~km/pixel is required to resolve such structures. 
This is consistent with the proposed resolution of $0.1R$ by \cite{2014ApJ...787L..22A}. Nevertheless, such fine structures are necessary but insufficient for the identification of KHI-induced turbulence.
%[Q: higher density more threads? or only caused by the higher velocity?]
%i.e., the loop width turns smaller, and the loop brightness is enhanced accordingly.
%shows the signatures of higher-order modes by the time evolution of the transverse intensity profile. Signatures: narrower loop width + more threads.

\subsubsection{Oscillatory pattern}
This subsection presents the oscillatory pattern extracted from the time--distance maps (see Appendix~\ref{append:B}) made from synthetic images. Loop centre positions were determined by Gaussian fits to the transverse intensity profiles, composing the oscillatory signals. The oscillatory behaviour of different loop segments forms a standing mode as expected, and oscillations across multiple LoS angles show amplitude variation, see more details in Appendix~\ref{append:B}. The latter is similar to previous synthetic observations \citep[e.g.,][]{2017ApJ...836..219A}, commonly due to the geometry effect.

\begin{figure}[ht]
	\centering
	\includegraphics[width=0.9\linewidth]{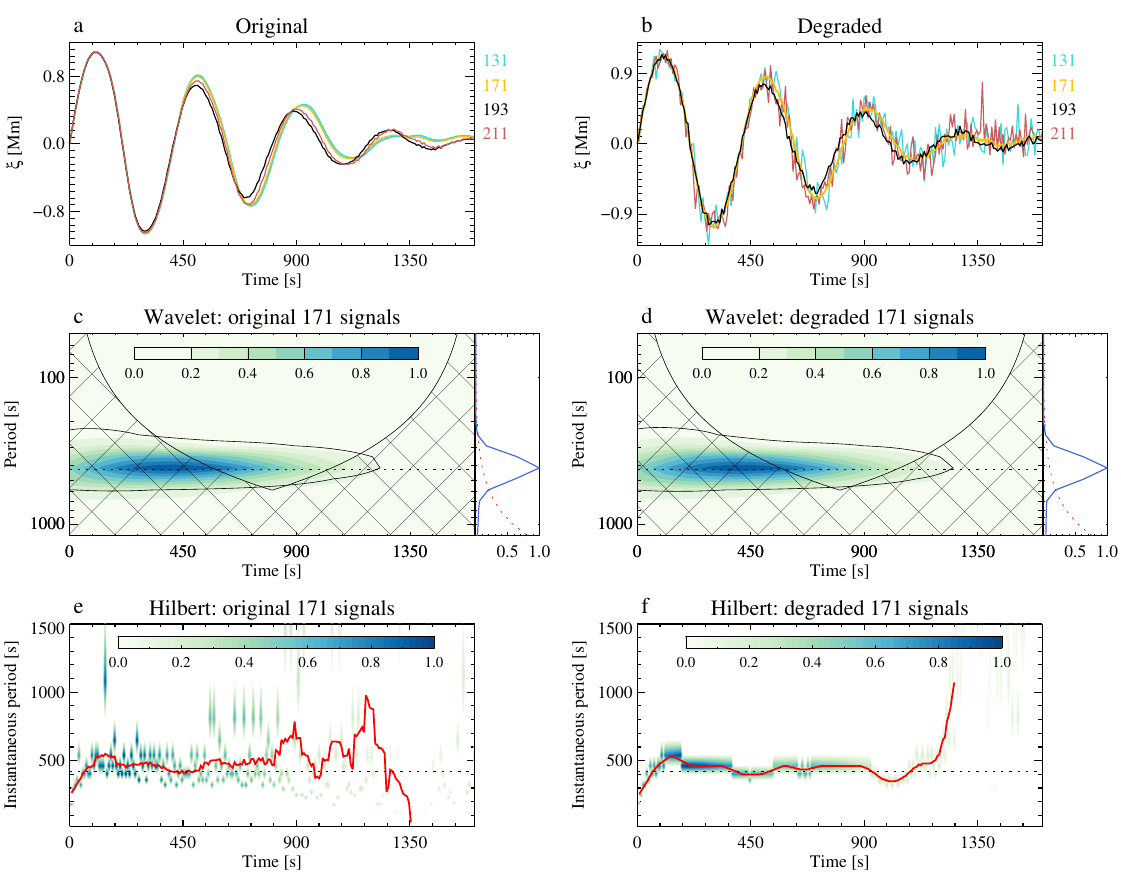}
	\caption{
        Oscillation signals of the loop with $\zeta=5, V_0=0.10$ extracted in different wavelengths and their time-frequency representation. Panels (a--b): oscillations in observed 131\,\AA, 171\,\AA, 193\,\AA, and 211 \,\AA\ in original simulation resolution (a) and AIA resolution (b), respectively. Panels (c--d): Wavelet Morlet spectrum for signals in 171\,\AA, with original resolution on the left, while the degraded versions are displayed on the right column. The black contour and the red lines in the right subpanel both represent the 95\% credible levels. The dotted horizontal lines denote the base period. The spectra amplitude is normalised by its maximum. Panels (e--f): Similar to (c--d) but for Hilbert spectrum. The red curves indicate the smoothed instantaneous periods. %Loop radius = 1 Mm.
	}
	\label{fig:spectra}
\end{figure}

The oscillating loop shows different properties in different wavelengths. In 171\,\AA, the loop brightness decreases and the loop width decreases over time, while in 193\,\AA, the loop brightnessis  enhanced and the loop becomes broader, see Appendix~\ref{append:B}.
A detailed comparison of oscillations across different wavelengths seen in the $0^{\circ}$ LoS, is presented in Fig.~\ref{fig:spectra} for a representative case with $\zeta=5, T_i/T_e=0.5$ and $V_0=0.10$. To study the impact of observational resolution, comparisons between the original and degraded AIA resolution are shown.
The overall pattern exhibits a frequency drift consistent with our KHI mixing model. This is confirmed by the Hilbert spectra in Fig.~\ref{fig:spectra}(e--f), see the increasing instantaneous period after $t=1000$. Unfortunately, Wavelet transformation is unable to capture such tiny frequency shift, due to the limited oscillation cycles \citep{2004SoPh..222..203D}. Though such a shift is small, this is a feature exactly predicted by our model (\citetalias{2024ApJ...966...68H,2025ApJ...991..208Z}).
Zooming in Fig.~\ref{fig:spectra}(a--b), signals in 131\,\AA\ almost overlap with those in 171\,\AA. In contrast, 193\,\AA\ and 211\,\AA\ signals have both slightly lower amplitudes (13.5\% and 6\% respectively) and a phase shift ($<\frac{\pi}{4}$) relative to 171\,\AA, suggesting faster apparent damping in hotter channels. In our turbulence damping theory, the damping rate of the oscillation depends on how much of the mixing layer participates in the collective motion, quantified by the density threshold $\rho_T$. When a larger portion of the mixing layer is included (corresponding to a smaller $\rho_T$), the decay becomes stronger, with reduced amplitude and an increased phase shift at the later stage. Whether the change of $\rho_T$ can explain oscillation difference across wavelengths will be evaluated in Section~\ref{sec:CoM_vs_CoE}. Similar phase behaviour was noted by \citet{2017ApJ...836..219A}, and is caused by each channel's temperature sensitivity: for $T_e>T_i$, 171\,\AA\ primarily traces the loop core, 193\,\AA\ and 211\,\AA\ the boundary. The expansion of the loop boundary due to turbulence makes oscillations recorded in boundary-tracing channels like 193\,\AA\ and 211\,\AA\ appear to damp faster. A higher $T_e$ (a lower $T_i/T_e$) produces a greater damping discrepancy between 171\,\AA\ and 193/211\,\AA, see $T_i/T_e=0.3$ in Fig.~\ref{fig:loop41_signals} in the Appendix~\ref{append:B}.
Differences in key parameters, including frequency/period, amplitude, and damping time, across wavelengths are typically below 50\%. Again, such a frequency shift is too tiny to be confidently detected by the time-frequency technique. Moreover, these discrepancies are blurred by noise at degraded resolution (Fig.~\ref{fig:spectra}b, also see Fig.~\ref{fig:loop4_signals}g--j). 
%Zhong\_loop4\_T has more threads than Zhong\_loop4.

\subsection{Centre of mass vs. centre of emission}
\label{sec:CoM_vs_CoE}
%starting with the $\rho^2$ question, moving to CoM–CoE threshold matching, then to temperature-based reasoning, and ending with turbulence-driven asymmetry.
In the above analysis, loop oscillations are described using CoM velocity and displacement in both analytic theory and simulation. Observationally, however, oscillations are traced via the Gaussian centre of the transverse intensity profile, i.e., CoE. We therefore test whether CoM motions can represent CoE displacement and thus whether our formula (Eq.(12) of \citetalias{2025ApJ...991..208Z}) can be applied to observed oscillations.
For simplification, we only focus on the CoE oscillation observed in the $0^{\circ}$ LoS which best records the oscillation.

\subsubsection{The weight of CoM}

\begin{figure}[ht]
	\centering
	\includegraphics[width=0.92\linewidth]{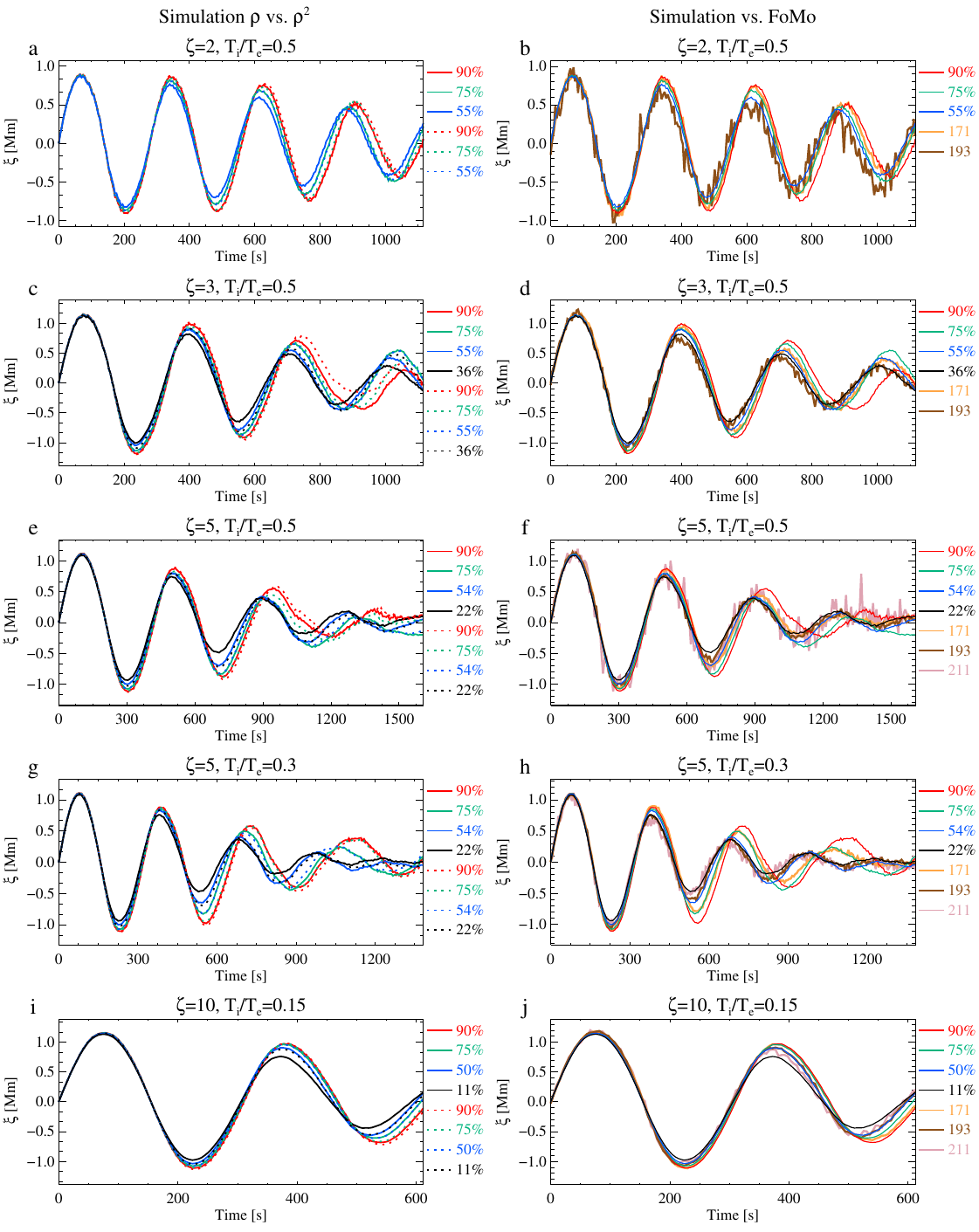}
	\caption{
        The comparison between the motion of the centre of mass or mass squared and that of the centre of emission, in different cases with varying density contrast and temperature contrast. First column: $\xi_{\rho\geq\rho_T}$ versus $\xi_{\rho^2\geq\rho_{T}^2}$. Second column: $\xi_{\rho>\rho_{T}}$ versus FoMo oscillations $\xi_\lambda$. %Third column: corresponding transverse intensity profile in multiple wavelengths at initial time (solid) and two cycles (dotted). For reference, the loop radius is 1~Mm. 
	}
	\label{fig:vs}
\end{figure}

Since EUV intensity is the LoS integral of $\rho^2$ (equivalent to $n_e^2$) weighted by the contribution function $G_{\lambda}(n_e,T)$, one might expect that computing the CoM with $\rho^2$ instead of $\rho$ would better match the CoE. However, as shown in Fig.~\ref{fig:vs} (first column), $\rho^2$-weighted CoM oscillations tracked in simulations are almost identical to $\rho$-weighted ones for $\rho_T\geq 90\%$, and differ at lower $\rho_T$, where $\rho^2$ weighting yields slightly slower decay. This effect is equivalent to comparing $\rho$-weighted CoM at a lower $\rho_T$ with $\rho^2$-weighted CoM at a higher $\rho_T$. Thus, $\rho^2$ weighting does not fundamentally change the displacement signal. 
%At lower $\rho_T$, CoM oscillations weighted by $\rho^2$ exhibit differences in amplitude and phase compared to those weighted by $\rho$. Specifically, the $\rho^2$-weighted oscillations tend to decay more slowly than their $\rho$-weighted counterparts at the same $\rho_T$. This is illustrated by that the former with a lower $\rho_T$ can match the latter with a higher $\rho_T$, e.g., $\xi_{\rho>0.75\rho_i}\sim\xi_{\rho^2>(0.9\rho_i)^2}$ and $\xi_{\rho>0.55\rho_i}\sim\xi_{\rho^2>(0.75\rho_i)^2}$. 
%Overall, CoM oscillations weighted by $\rho^2$ are qualitatively similar to those weighted by $\rho$, either at the same or a higher density threshold, suggesting that $\rho^2$ weighting does not provide substantially new information in this context.

In general, both CoM and CoE exhibit the same base period and initial amplitude, indicating that the key nonlinear features, including (1) period increase and (2) reduced amplitude, are preserved in synthetic imaging data. 
    Comparing CoM motions above various $\rho_T$, $\xi_{\rho\geq\rho_T}$, with CoE displacements extracted from synthetic images (Fig.~\ref{fig:vs}, second column) shows that CoE oscillations typically decay faster and exhibit earlier, stronger frequency drift than core motions (see the red curve representing $\xi_{\rho\geq90\%\rho_i}$). Matches can occur for specific $\rho_T$ values, e.g., 171\,\AA\ with $\rho_T=90\%\rho_i$ (gold vs. red in Fig.~\ref{fig:vs}b), 193\,\AA\ with $\rho_T=36\%\rho_i$ (brown vs. black in Fig.~\ref{fig:vs}d), and 211\,\AA\ with $\rho_T=50\%\rho_i$ (pink vs. blue in Fig.~\ref{fig:vs}j). But 171\,\AA\ sometimes traces plasma down to $\sim0.5\rho_i$ due to its response peaking near 0.6~MK and remaining significant in boundary regions. Overall, hotter channels tend to correspond to lower $\rho_T$.

\subsubsection{Density-threshold-dependent CoM}\label{sec:rho_T}
The above analysis demonstrates that the CoE oscillations in different wavelengths can be approximated by CoM motions above different $\rho_T$ in simulations. Considering the seismological application in observation, do oscillations in a specific wavelength generally correspond to a certain $\rho_T$? Concerning validating our turbulence damping model, several questions are naturally raised: Can our CoM-based formula explain CoE oscillations? If not, what are the possible reasons? Can we adjust the weight of density appropriately so that the CoM match the CoE? Before designing analyses to answer these questions, we should keep in mind the general discrepancies between the simulations and the model, as well as their impact. Section~\ref{sec:result_simulation} shows that the existence of higher-order modes makes the CoM/$V_{\rm CoM}$ oscillation amplitudes smaller than theoretically expected. Thus, the input $V_i$ in the model should be smaller than the measured value in simulations. In addition, when the kinetic energy of the core approaches that of the mixing layer, the decay observed in the simulation is faster than the model. The faster decay manifests as (1) the time to start dramatic decay (amplitude drops and phase shifts) arrives earlier; (2) the time when the loop becomes fully turbulent arrives earlier. To match the stronger damping in the later stage while keeping the same in the earlier stage, the model should have a systematically greater $\zeta$ (increase within 1, the exact value also depends on other parameters), or a combination of smaller $\rho_T$, a smaller $C_1/R$, and a greater $\zeta$. And this is confirmed by the results of fitting the simulated oscillations with the model (Section~\ref{sec:model_fit}). A smaller $\rho_T$ causes faster overall decay, which could be compensated for by the latter two. These impacts will naturally be inherited in the comparison between the CoE oscillation and the model.

\begin{figure}[ht]
	\centering
	\includegraphics[width=0.9\linewidth]{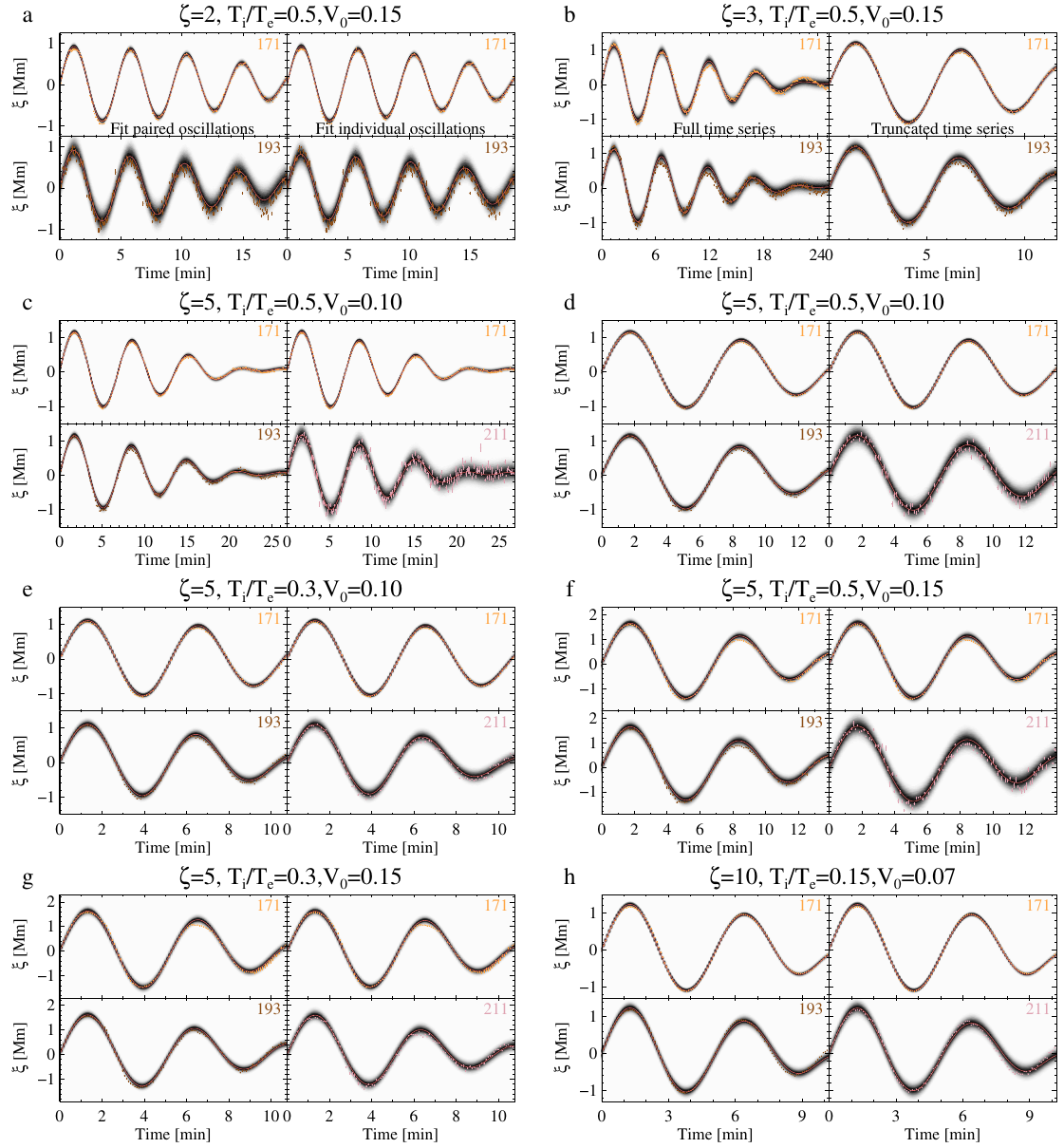}
	\caption{
        Fitting the CoE oscillations in different AIA channels with the nonlinear turbulence damping function. The input oscillation signals are denoted by dots with error bars. The best fits are denoted by the red curves, and the predictive posterior distribution is represented by the grey shading. (a) The comparison between fitting the paired oscillations and fitting the individual oscillations. Fits in the remaining panels are all obtained by fitting with paired functions, with oscillations in two channels displayed in parallel in vertical directions. (b) Full time-series vs. truncated by the time where the kinetic energy of the core equals that of the mixing layer. Similar for (c--d) but for case with $\zeta=5,T_e=2[\rm{MK}],V_0=0.10$.
	}
	\label{fig:fits}
\end{figure}

To systematically check if a specific wavelength can be related to a certain $\rho_T$, we fit the CoE oscillation using our nonlinear function. [This function translates the role of $\rho_T$ to $\eta$, representing the fraction of total mass per unit width between the core and the mixing layer (\citetalias{2025ApJ...991..208Z}).] %We assume 171\,\AA\ captures higher-density plasma near $\rho_i$, while 193\,\AA\ and 211\,\AA\ reflect lower-density plasma closer to $\rho_e$. 
We apply SoBAT to fit pairs of CoE oscillations simultaneously in different channels using the following model:
\begin{align}\label{eq:fomo_fit}
\begin{cases}
    \xi_{\lambda_1} &= F(C_1/R, \zeta,V_i,{\rm P_k},\eta_{1}),\\
    \xi_{\lambda_2} &= F(C_1/R, \zeta,V_i,{\rm P_k},\eta_{2}=\alpha\eta_{1}).
\end{cases}
\end{align} 
\noindent where $\alpha$ is the ratio $\eta_{1}/\eta_{2}$, where the subscripts corresponds to the wavelength $\lambda_{1/2}$. 
Fitting paired oscillations and fitting individual oscillations yield similar outcomes, including the posterior distributions, see the comparison between the left and right columns in Fig.~\ref{fig:fits}a. But the posterior distribution in 193\,\AA\ in the case of pair fit (see the grey shading on the left of Fig.~\ref{fig:fits}a) does not cover the last trough, while the individual fit does, this is because the input of paired oscillations contains more information and thus puts more constraint on the parameters. We first consider fitting the full time series. Then, taking into account that the theory assumption does not hold hence decreasing the model accuracy in the later stage (\citetalias{2024ApJ...966...68H}), we truncate the oscillations at when the mass of the loop core stops linear loss (labelled by \lq\lq T1\rq\rq) and at when the kinetic energy of the loop core equals that of the mixing layer (labelled by \lq\lq T2\rq\rq) and re-run the fitting. The fitting results are displayed in Fig.~\ref{fig:fits} and Table~\ref{tab:fit}. For the fitted $\eta$, we estimate $\rho_{T}$ using the average density profile for the corresponding density contrast, see the results listed in Table\ref{tab:fit} for all analysed cases. Here $\rho_T$ is normalised as $\frac{\rho_T-\rho_e}{\rho_i-\rho_e}$ to reveal its relative position between $\rho_i$ and $\rho_e$: a value of 1 (0) corresponds to $\rho_T=\rho_i$ ($\rho_T=\rho_e$). 

For pair fit results, the posterior distribution represented by the grey shadings in Fig.~\ref{fig:fits} generally covers well the input signals (see dots with error bars). However, comparing the best-fitted values and the measured values as listed in Table~\ref{tab:fit}, the fits only recover some key simulation parameters within errors, such as $V_i$ and $\rm{P_k}$. In the model, $V_i$ controls the initial displacement amplitude and $\rm{P_k}$ decides the periodicity of the oscillation. Fits with the full time series gives either a smaller $C_1/R$ or a greater density contrast $\zeta$, except that case with $\zeta=10$ has both greater values for $C_1/R$ and $\zeta$. {In addition, 7 out of 9 fits give a $\rho_{T,171}$ greater than 0.5$\rho_i$ and only 2 are greater than 0.8$\rho_i$, meaning that 171\,\AA\ targets both the core and part of the mixing layer. From the perspective of the role of these parameters in decay, an increment of $\zeta$ within 1 causes only slightly enhanced decay in the last cycle with a bigger phase drift, and the brake of decay caused by the smaller $C_1/R$ can be compensated by the decrease of $\rho_T$. In some cases, e.g., $\zeta=5,V_i=15$, the CoE oscillation exhibits a distorted waveform in the last cycle, as it includes signals in the fully turbulence phase, thus the best fit forces the $\rho_T$ approach zero. This problem can be avoided by truncating the oscillations.
Example comparisons between fitting the full time series and fitting the truncated time series are shown in Fig.~\ref{fig:fits}b and c--d for the case of $\zeta=3$ and $\zeta=5$, respectively. From visual inspection, removal of signals in the later stage makes no big difference in the posterior distributions. Comparing the parameter distributions in Table~\ref{tab:fit}, the shorter the input segment gives the greater $C_1/R$, the greater $\zeta$, and the greater $\rho_T$, reconstructing less decay. Among all, only the case with $\zeta=5, T_e=2$~[MK] and T2 gives posteriors that agree with simulation-measured values. Most cases give a greater $\zeta$, possibly to match the greater phase shift.}
%[A greater $C_1/R$ can be compensated for a greater $\zeta$] 
%An example of the posterior distribution of parameters is shown in Fig.~\ref{fig:corner}, where all parameters are well constrained and agree with simulation-derived values. 
 
Concerning the relationship between $\rho_T$ and wavelengths. As expected, 171\,\AA\ oscillations generally correspond to relatively higher $\rho_T$ than hotter channels. For 171\,\AA, all the truncated cases gives $\frac{\rho_T-\rho_e}{\rho_i-\rho_e}\geq 0.5$ and more than half is greater than 0.6, indicating $\rho_{T,171}$ is closer to $\rho_i$ rather than $\rho_e$. Most 193\,\AA\ signals are associated with low $\rho_T$ approaching $\rho_e$. The 211\,\AA\ results are less consistent, likely due to low signal-to-noise ratio and limited sampling.
An exception occurs when the $T_e$ is high. For example, in cases with $T_e=6.7$~MK or $T_e=3.3$~MK, all three channels are associated with high $\rho_T$, in descending order from 171\,\AA\ to 211\,\AA. In this case, the oscillations in different channels only have minor differences. One possible reason is that the size of the mixing layer is not great enough to make a contribution in hotter channels.

%%\textcolor{red}{Truncate the oscillations by the time when the kinetic energy of the core equals that of the mixing layer for fitting gives a greater $C_1/R$, a greater $\zeta$, and a greater $\rho_T$. But if the time duration is too short, the oscillations in different channels are virtually the same, as the mixing layer is not developed enough, and its width is small. It is the contribution of the mixing layer that makes the oscillations in multiple wavelengths differentiate. [$V_i$ determines the amplitude? $C_1/R, V_i, \rho_i,\rho_T$ jointly decide the decay.][Put in discussion?]}

\begin{figure}[t]
	\centering
    \includegraphics[width=\linewidth]{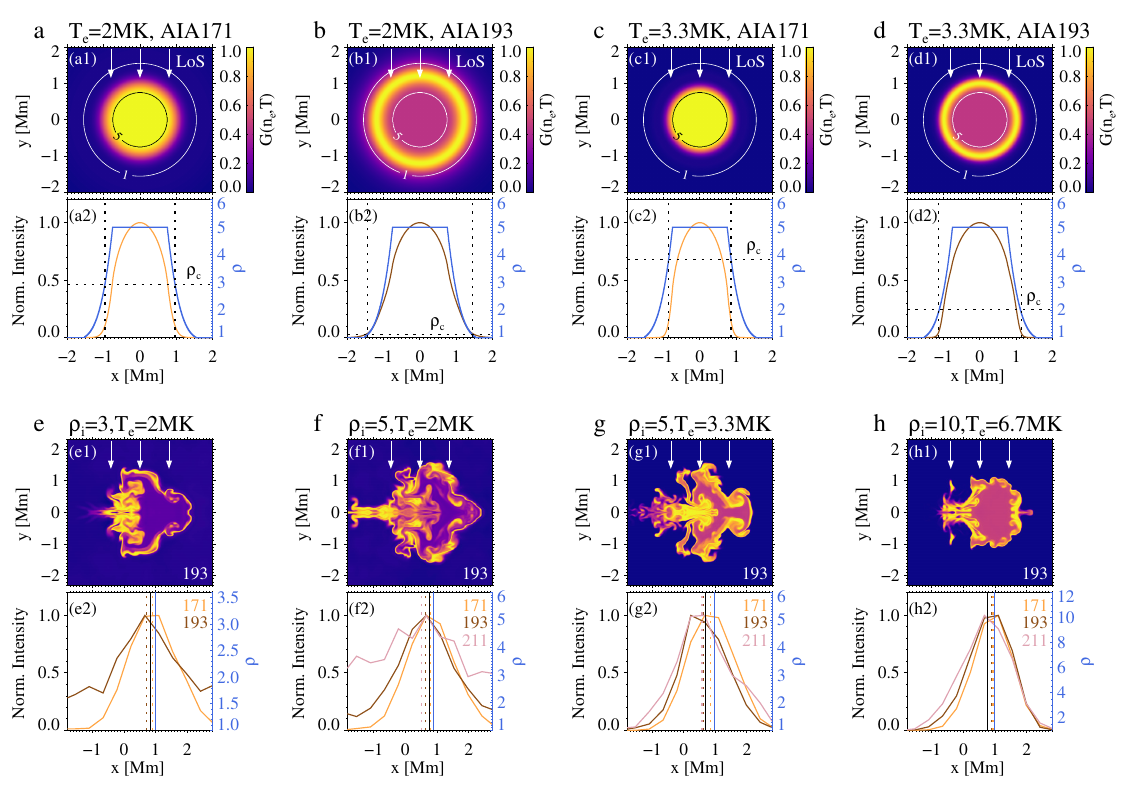}
	\caption{
        Contribution function of the loop's cross-section (panels labelled by 1) and corresponding transverse density and intensity profile (panels labelled by 2) in AIA 171\,\AA\ and 193\,\AA\ seen in the perpendicular ($0^{\circ}$) LoS (indicated by the arrows). Panels (a--d) are for ideal estimations of the loops with $\rho_i=5$ using the KHI mixing theory. The average density and temperature are calculated when the mixing/turbulent layer reaches a height of 0.8$R$, before being input into the computation of the contribution function. The contours in the first row denote the density of 5 and 1, and the white arrows indicate the LoS direction. In the second row, the vertical lines indicate the location of maximum gradient of contribution across the profile, and the horizontal lines point out the corresponding density threshold $\rho_c$. Panels (e--h) are for actual measurements in simulations and synthetic AIA images for cases with different $\rho_i$ and $T_e$. Panels (e1--h1) are the contribution of the loop's cross-section in 193\,\AA\ at the second oscillation crest. The vertical dotted lines in panels e2--h2 indicate the CoE displacement obtained by Gaussian fitting in different wavelengths. The blue and black vertical solid lines represent the CoM$_{\rho>1.1}$ and CoM$_{\rho>0.9\rho_i}$ respectively.
	}
	\label{fig:coe}
\end{figure}

%While the loop contains plasma spanning a range of densities and temperatures, the observed emission in each wavelength primarily reflects plasma within the temperature range where the instrument's response function is significant (see Fig.~\ref{fig:emission}). Now we consider temperature constraints based solely on the response function. 
The descending order of $\rho_{T,171}>\rho_{T,193}>\rho_{T,211}$ can be explained by the emissivity response structure. %Temperature structure offers a first-order explanation 
Recall Fig.~\ref{fig:emission}, the emissivity in each wavelength primarily reflects plasma within the temperature range where the contribution function is significant.
As shown in Fig.\ref{fig:coe}a and c, for a loop cross-section with a turbulent layer of 0.8$R$ in width, the 171\,\AA\ response decreases monotonically across the radius (see panels a1 and c1), qualitatively synchronises with the density profile, producing a sharp intensity profile in the $0^{\circ}$ LoS (see panels a2 and c2). In contrast, 193\,\AA\ response peaks in a torus surrounding the inner core (Fig.\ref{fig:coe}c1 and e1), making this channel and other hotter wavelengths more sensitive to boundary dynamics. Because their response does not follow the monotonic density decrease, their intensity profiles are broader and flatter (Fig.\ref{fig:coe}c2 and e2). %From inside out the loop, they do not synchronise with the decreasing density profile, leading to a flatter and broader intensity profile
%The intensity profile covers the plasma above $\rho_T$ where $\rho_T$ is the lower limit corresponding to the boundary of the intensity profile. Therefore, the CoE can be estimated by CoM with $\rho>\rho_T$. %The emission is highly dependent on the response which is mainly based on temperature distribution.
Therefore, loop emission (which is higher than the background) bounds at a temperature/density threshold where response falls off, and such a threshold for 171/193/211\,\AA\ is decreasing. But the density threshold determined by where the response drops at its maximum gradient does not always agree with the fitted $\rho_T$, indicating that temperature constraints alone can not fully explain CoE behaviour.
%[[REMOVE]To estimate this threshold from thermal structure, we define the upper bound of temperature in the high-response region \textcolor{red}{at its maximum gradient} as $T_{c,\lambda}$: $1.2$~MK for 171\,\AA, $1.9$~MK for 193\,\AA, and $2.2$~MK for 211\,\AA. \textcolor{red}{Then we infer $\rho_{c,\lambda}$ at the location of the temperature cutoff $T_{c,\lambda}$} according to that the ideal gas law relates the transverse profile of temperature and density. For example, in the case of $\zeta=5,T_i/T_e=0.3$, inversion yields $\rho_{c,171}=3.5, \rho_{c,193}=2.0, \rho_{c,211}=1.65$. Increasing $T_e$ or $\rho_i$ results in a narrower high-response region, a sharper intensity profile, and a higher $\rho_c$.  The normalised values $\frac{\rho_c-\rho_e}{\rho_i-\rho_e}$ for each channel are shown in Table\ref{tab:2}. Comparing these values with the fitted results $\frac{\rho_T-\rho_e}{\rho_i-\rho_e}$, we find good agreement in 5 out of 8 cases for 171\,\AA, but only 3 for 193\,\AA\ and none for 211\,\AA. This suggests that temperature constraints alone can not fully explain CoE behaviour.]

%[EUV emission depends on both the squared plasma density and the temperature-density-dependent response function. Therefore, the apparent motion of the loop, tracked via its emission centre, does not directly reflect either the CoM or a purely temperature-weighted centroid. The observed signal results from a coupled effect of density, temperature, and LoS integration. These factors must be jointly considered in future efforts to derive an analytical expression for CoE displacement.]

Ideally, given the loop’s initially circular cross-section and symmetric density and temperature distribution, the CoE at a given wavelength coincides with the CoM$_{\rho\geq\rho_T}$. Early in the evolution, this is indeed the case (see Fig.~\ref{fig:vs}). 
However, as turbulence develops and penetrates inward, the discrepancies occur. 
The turbulence and higher-order modes deform the loop cross-section, breaking symmetry and producing head–tail asymmetry. As shown in Fig.~\ref{fig:coe}e1--h1, the dense core compresses while the low-density, high-temperature boundary fragments. Consequently, 171\,\AA\ emission remains core-centred, while hotter channels like 193\,\AA\ weigh more toward the tail, giving systematically smaller CoE displacements than CoM$_{\rho\geq\rho_T}$ (see vertical lines in Fig.~\ref{fig:coe}e2--g2). A similar outcome is reported in a synthetic EUV observation of flux rope kink instability \citep{2017ApJ...842...16S}. %As a result, the peak of the 171\,\AA\ intensity profile may deviate slightly from the CoM, whereas hotter channels shift further toward the tail.
%This leads to systematically smaller CoE displacements than the CoM above the predicted $\rho_c$, and CoE displacements in hotter channels are smaller than those in 171\,\AA\ (see vertical lines in Fig.~\ref{fig:coe}e2--h2).
In denser loops with the same nonlinearity, turbulence grows more slowly, preserving cross-section coherence and reducing tail bias. In such cases (e.g., Fig.~\ref{fig:coe}h2), CoE displacement in hotter channels can exceed those of CoM$_{\rho\geq\rho_T}$.
We conclude that the divergence between CoE and CoM primarily arises from nonlinear cross-section deformation, with emissivity response and turbulence jointly controlling the wavelength-dependent displacement bias.
As a result, at an early stage where the turbulence is not well developed, oscillation of CoE$_{\lambda}$ nearly matches that of CoM$_{\rho\geq\rho_{T,\lambda}}$, as the turbulence layer grows to a certain size, CoE$_{\lambda}$ displacement is lower than that of CoM$_{\rho\geq\rho_{T,\lambda}}$, even with a phase shift. To reproduce such enhanced decay at the later stage while keeping the beginning the same in CoM oscillation, the turbulence damping model must have a greater $\zeta$ and a smaller $\rho_T$.

\begin{table}
    \centering
    \begin{tabular}{cccccccccccc}
    \hline
    \multicolumn{2}{l}{Simulation parameters} & & & \multicolumn{7}{c}{Fitted parameters} \\ \cline{0-1} \cline{5-11} 
     $V_0(\frac{V_0L}{C_{\rm k}R})$ & $T_e [\rm{MK}]$ & & Seg & $C_1/R$ & $\zeta$ & $V_i$~[km~s$^{-1}$] & $\rm{P_k}$~[s] & $\frac{\rho_{T,171}-\rho_e}{\rho_i-\rho_e}$ &  $\frac{\rho_{T,193}-\rho_e}{\rho_i-\rho_e}$ & $\frac{\rho_{T,211}-\rho_e}{\rho_i-\rho_e}$ \\
      \hline\hline
     0.15 (4.50) & 2 & Sim. values & & $0.31^{+0.04}_{-0.03}$ & 2 & $24\pm4$ & 279  \\
      &  & Fitted values  & Full & $0.30^{+0.01}_{-0.01}$ & $3.2^{+0.1}_{-0.1}$ & $20^{+1}_{-1}$ & $280^{+1}_{-1}$  & $0.67^{+0.12}_{-0.11}$ & $0^{+0.12}_{-0.01}$ & - \\ \hline
     0.15 (4.53) & 1.25 & Sim. values & & $0.30^{+0.04}_{-0.04}$ & 2 & $19\pm3$ & 353  \\
       & & Fitted values & Full & $0.33^{+0.02}_{-0.03}$ & $3.5^{+0.2}_{-0.2}$ & $16^{+1}_{-1}$ & $355^{+2}_{-2}$  & $0.64^{+0.15}_{-0.18}$ & - & - \\ \hline
     0.15 (5.23) & 2 & Sim. values &  & $0.39^{+0.08}_{-0.05}$ & 3 & $26\pm4$ & 327  \\
     & & Fitted values & Full & $0.27^{+0.01}_{-0.01}$ & $4.2^{+0.2}_{-0.1}$ & $22^{+1}_{-1}$ & $326^{+1}_{-1}$ & $0.29^{+0.10}_{-0.07}$ & $0^{+0.01}_{-0.00}$ & - \\
      &     &   & T1 & $0.36^{+0.01}_{-0.01}$ & $4.2^{+0.2}_{-0.2}$ & $23^{+0}_{-0}$ & $329^{+1}_{-1}$ & $0.49^{+0.13}_{-0.08}$ & $0^{+0.06}_{-0}$ & - \\
      &     &   & T2 & $0.38^{+0.01}_{-0.03}$ & $5.1^{+0.2}_{-0.3}$ & $24^{+0}_{-0}$ & $329^{+1}_{-1}$ & $0.57^{+0.20}_{-0.14}$ & $0.07^{+0.19}_{-0.07}$ & - \\ \hline
    0.10 (3.92) & 2 & Sim. values & & $0.45^{+0.08}_{-0.06}$ & 4 & $17\pm2$ & 372  \\
    & & Fitted values & Full &$0.40^{+0.03}_{-0.01}$ & $4.8^{+0.1}_{-0.1}$ & $17^{+0}_{-0}$ & $373^{+1}_{-1}$ & $0.41^{+0.16}_{-0.10}$ &  $0.21^{+0.16}_{-0.02}$ & $0.25^{+0.19}_{-0.23}$ \\
    & &  & T2 &$0.44^{+0.07}_{-0.05}$ & $5.5^{+0.4}_{-0.4}$ & $17^{+0}_{-0}$ & $372^{+1}_{-1}$ & $0.49^{+0.22}_{-0.21}$ &  $0.12^{+0.22}_{-0.11}$ & $0.24^{+0.35}_{-0.23}$ \\ \hline
    0.10 (4.27) & 2 & Sim. values & & $0.49^{+0.11}_{-0.08}$ & 5 & $17\pm2$ & 422  \\
    & & Fitted values & Full &$0.37^{+0.00}_{-0.01}$ & $4.9^{+0.1}_{-0.1}$ & $17^{+0}_{-0}$ & $419^{+1}_{-1}$ & $0.26^{+0.06}_{-0.05}$ &  $0.00^{+0.02}_{-0.00}$ &  $0.16^{+0.08}_{-0.12}$ \\
    &  &  & T2 &$0.48^{+0.05}_{-0.05}$ & $5.5^{+0.3}_{-0.3}$ & $17^{+0}_{-0}$ & $418^{+1}_{-1}$ & $0.59^{+0.16}_{-0.18}$ &  $0.31^{+0.16}_{-0.19}$ &  $0.43^{+0.21}_{-0.28}$ \\ \hline
    0.10 (4.33) & 3.3 & Sim. values & & $0.62^{+0.13}_{-0.10}$ & 5 & $22\pm3$ & 323 \\
    & & Fitted values & Full &$0.37^{+0.01}_{-0.01}$ & $5.1^{+0.1}_{-0.1}$ & $22^{+0}_{-0}$ & $321^{+1}_{-1}$ & $0.46^{+0.06}_{-0.05}$ &  $0.00^{+0.02}_{-0.19}$ &  $0.00^{+0.01}_{-0.00}$ \\
    & &   & T1 &$0.56^{+0.03}_{-0.03}$ & $5.7^{+0.4}_{-0.4}$ & $22^{+0}_{-0}$ & $319^{+1}_{-1}$ & $0.82^{+0.11}_{-0.12}$ & $0.51^{+0.11}_{-0.13}$ & $0.37^{+0.12}_{-0.12}$\\
     &   &  & T2 &$0.54^{+0.07}_{-0.07}$ & $5.8^{+0.5}_{-0.4}$ & $22^{+0}_{-0}$ & $318^{+1}_{-1}$ & $0.81^{+0.15}_{-0.17}$ & $0.47^{+0.19}_{-0.21}$ & $0.36^{+0.20}_{-0.22}$\\ \hline
    0.15 (6.50) & 2 & Sim. values & & $0.54^{+0.17}_{-0.10}$ & 5 & $26\pm6$ & 422  \\
    & & Fitted values   & Full &$0.31^{+0.01}_{-0.01}$ & $5.3^{+0.1}_{-0.2}$ & $26^{+0}_{-1}$ & $419^{+1}_{-3}$ & $0.00^{+0.04}_{-0.00}$ &  $0.00^{+0.01}_{-0.00}$ &  $0.00^{+0.12}_{-0.00}$ \\
      & &  & T1 & $0.45^{+0.01}_{-0.02}$ & $4.7^{+0.2}_{-0.2}$ & $25^{+0}_{-0}$ & $423^{+2}_{-2}$ & $0.47^{+0.16}_{-0.19}$ &  $0.30^{+0.12}_{-0.10}$ &  $0.37^{+0.13}_{-0.18}$ \\
      &  &   & T2 & $0.47^{+0.01}_{-0.02}$ & $4.8^{+0.3}_{-0.2}$ & $25^{+1}_{-0}$ & $421^{+2}_{-2}$ & $0.55^{+0.15}_{-0.18}$ &  $0.40^{+0.11}_{-0.13}$ & $0.48^{+0.12}_{-0.19}$ \\ \hline
      0.15 (6.40) & 3.3 & Sim. values & & $0.58^{+0.16}_{-0.10}$ & 5 & $36\pm8$ & 323  \\
      & & Fitted values   & Full &$0.30^{+0.01}_{-0.01}$ & $4.9^{+0.1}_{-0.1}$ & $32^{+0}_{-1}$ & $321^{+2}_{-2}$ & $0.19^{+0.08}_{-0.08}$ &  $0.00^{+0.01}_{-0.00}$ &  $0.00^{+0.01}_{-0.00}$ \\
      & & & T1 & $0.38^{+0.03}_{-0.01}$ & $5.0^{+0.2}_{-0.2}$ & $32^{+0}_{-0}$ & $322^{+0}_{-2}$ & $0.57^{+0.15}_{-0.09}$ &  $0.02^{+0.11}_{-0.02}$ &  $0.00^{+0.05}_{-0.00}$ \\
       &   &   & T2 & $0.43^{+0.02}_{-0.03}$ & $5.1^{+0.2}_{-0.2}$ & $32^{+0}_{-0}$ & $321^{+2}_{-2}$ & $0.72^{+0.12}_{-0.16}$& $0.25^{+0.15}_{-0.17}$ & $0.04^{+0.17}_{-0.04}$ \\ \hline
     0.07 (4.04) & 6.7 & Sim. values & & $0.75^{+0.13}_{-0.10}$ & 10 & $27\pm4$ & 309  \\
     & & Fitted values & Full & $0.96^{+0.03}_{-0.04}$ & $12.8^{+0.7}_{-0.5}$ & $25^{+0}_{-0}$ & $308^{+1}_{-1}$ & $0.82^{+0.07}_{-0.09}$ &  $0.74^{+0.06}_{-0.08}$ & $0.69^{+0.07}_{-0.10}$ \\
       \hline
    \end{tabular}
    \caption{The best fitted parameters for oscillations in multiple wavelengths fitted with the turbulence damping function (Eq~\ref{eq:fomo_fit}). Uncertainties are reported as 95\% credible intervals of the posterior distribution. For each case, the first row lists the setup conditions ($T_e, V_0 (\frac{V_0L}{C_{\rm k}R}),\zeta$) and measured values ($C_1/R,V_i,\rm{P_k}$) in the corresponding simulation, while the remaining rows display the fitted values. The internal temperature of the loops is 1~MK. The column of \lq\lq Seg\rq\rq, denotes the selected time segment of the input oscillation, with \lq\lq Full\rq\rq\ indicate the full time series, \lq\lq T1\rq\rq\ for truncating the oscillations at when the core mass stops linear loss and \lq\lq T2\rq\rq\ for truncating the oscillations at when the kinetic energy of the loop core equals that of the mixing layer. Note the case with $\zeta=2,T_e=1.25$~[MK], only oscillation in 171\,\AA\ is visible so it is fitted by a single function rather than paired functions as the remaining cases.}
    \label{tab:fit}
\end{table}

\section{Discussion}\label{sec:discussion}

\subsection{Period increase}
\label{sec:dis_period}

%The perturbation set in this work ($\xi_0\geq R$) is much stronger than the previously existing simulations \citep[e.g., $\xi_0\leq0.5R$ in][]{2016A&A...595A..81M,2019FrP.....7...85A}.  %say, $\xi_0=0.5R$ in \citet{2016A&A...595A..81M}, $\xi_0=0.44R$ in [Antolin+2019,ApJL], $A_0L/R=3.3$ in \citet{2019FrP.....7...85A}. v0L/C_{\rm k}R=1.9 in Antolin+2017. [use a common measure for comparison.] %0.05 in roi2 = 0.5R.
Statistically, for nonlinear oscillation with $\frac{V_0L}{C_{\rm k}R}> 2$, the oscillation period increases by a few percent relative to the linear kink period, with the increase scaling with amplitude. This behaviour has not been reported in previous simulations of nonlinear kink oscillations, which employed smaller perturbations and weaker nonlinearity \citep[e.g., $\xi_0\leq0.5R$ in][]{2016A&A...595A..81M,2019FrP.....7...85A} compared to the present work ($\xi_0\geq R$).

Several factors could contribute to this period increase. 
One possible mechanism is that the periodic boundary conditions in our simulation domain make the loop feel copies of itself. This idea is motivated by studies in damped kink oscillations of two interacting loops \citep[e.g.,][]{2015A&A...582A.120S,2024A&A...686A...2S}, where the fundamental kink period increases with the separation between two loops $d$.
In our setup, with $R=0.5, d=5.4$ for a periodic boundary, Eq(30) of \citet{2015A&A...582A.120S}, $P = P_{\rm k}\sqrt{1\mp\frac{\zeta-1}{\zeta+1}(\frac{R}{d})^2}$ predicts a period increase $<0.4\%$, far smaller than the $\sim1-5\%$ increase measured in our simulations. Accounting for two periodic boundaries effectively produces a system of three interacting loops. In this three-identical-loop configuration, \citet{2009ApJ...692.1582L} showed that the central loop exhibits fluting-like motions with negligible transverse displacement, and kink-like motions with the lowest frequency $3.6\%$ smaller than the linear kink frequency for $\zeta=3$ and the separation distance of $3R$. Assuming the frequency drop scales with $d^{-2}$ as in the two-loop configuration, for a larger separation distance in our work, it is around 1\%.
%Since we have two periodic boundaries, we can treat the system as three interacting loops, but this results in the middle loop exhibiting fluting-like motions (no transverse displacement), according to \citet{2009ApJ...692.1582L}. 

%Geometrical perspectives.
Loop geometry may also influence the oscillation period. 
In the linear regime, a straight loop with an elliptical cross-section oscillating along its major axis shows an increased period with increasing aspect ratio $a/b$, approaching $\frac{2L}{C_{Ai}}$ as $a/b\to \infty$ \citep{2003A&A...409..287R,2020ApJ...904..116G,2022MNRAS.514.4329L}. This could be relevant here, as our loop becomes significantly deformed and departs from a circular cross-section. 
For curved loops with magnetic field stratification but uniform density, vertically polarised kink oscillations have been shown to exhibit periods $\sim9\%$ longer than the WKB-predicted kink period \citep{2024A&A...687A..30G}, although such deviations are within observational uncertainties and do not significantly affect MHD seismology. In density-stratified loops with a density contrast of 10 and scale height $>2$, nonlinear kink periods also increase by a few percent \citep{2014SoPh..289.1999R}. 
In our simulations, the loop is initially straight with uniform density along its length, but large-amplitude perturbations introduce some curvature. A bent loop has a longer effective length, and hence a longer period. Estimating the length increase as $\sqrt{L_z^2+\xi_0^2}/L_z$ gives $\sim1\%$ for $\xi_0=R$, still smaller than the measured increase.
We therefore conclude that the observed period increase is primarily a manifestation of a nonlinear effect.
%could be that the loop cross-section deformation/asymmetry causes CoM to deviate from that estimated under a circular cross-section assumption.

\subsection{Applicability of the nonlinear turbulence-damping model}
\label{sec:dis_model}
Our model explains how KHI-induced turbulence develops in the oscillating loops and modifies the oscillation damping profile. It performs well in several aspects.
% Evidence showing our model works:
As shown in the supplementary animation, the average density profile measured from the simulation agrees closely with the theoretical prediction until the kinetic energy in the boundary layer exceeds that of the loop core. When the boundary kinetic energy dominates, it has a back interaction with the core, so the direction of momentum transfer no longer follows the core-to-boundary flow assumed in the model \citep{2024ApJ...966...68H}.
The nonlinear damping functions we derived for CoM velocity and displacement closely match those measured in simulations, capturing features such as the amplitude decay and velocity evolution, though not the period increase (Fig.~\ref{fig:simul_curve}b) or amplitude reduction (Fig.~\ref{fig:simul_curve}d). 
Furthermore, applying our nonlinear model to fit oscillations observed in synthetic 171\,\AA\ images successfully recovers key physical parameters consistent with those from the simulation.

Despite the good overall agreement between the simulated transverse loop oscillations and our model, the posterior distributions of the fitted parameters reveal that the degeneracy exists among the mixing efficiency $C_1$, density contrast $\zeta$, and mixing layer participation $\eta~(\rho_T)$, as shown in Section~\ref{sec:model_fit}. Specifically, $C_1$ is positively correlated with $\zeta$ and negatively correlated with $\eta$ (equivalently, positively correlated with $\rho_T$). These three parameters jointly determine the damping profile. In contrast, the remaining two parameters, initial velocity amplitude $V_i$ and effective oscillation period $\rm{P_k}$, are well-constrained. From a seismological perspective, $\rm{P_k}$ and $V_i$ can be robustly recovered via fitting observations with our model, while other parameters cannot be uniquely determined. A further observational constraint on at least one of these parameters can solve the issue. For example, the density contrast can be determined by intensity contrast for a rough estimation, or by the Differential Emission Measure for a more precise estimation.

The consistency between the model and simulations holds within the regime where its assumptions remain valid; beyond this regime, the model exhibits expected limitations. First, the model describes average loop behaviour and is not designed to capture local or fine-scale dynamics. %It also neglects the observational view angle, which may alter the observed signal. [just a linear transformation]
Second, it is designed primarily for idealised conditions where turbulence generated by KHI dominates the damping. It does not account for other damping mechanisms or nonlinear phenomena such as the generation and cascade of higher-order modes, nor does it capture the observed period increase. 
Consequently, its application is restricted to regimes of weak and moderate nonlinearity, where KHI-driven turbulence dominates and higher-order mode interactions remain relatively weak. Beyond this regime, additional physical effects must be incorporated.
Recently, \citet{2019FrP.....7...85A} showed that for $\frac{V_0L}{C_{\rm k}R}=2.6$, $m> 2$ modes dominate at the beginning of oscillation, and an $m=2$ mode is formed alongside KHI, reaching power comparable to $m=1$ at some instances of time. And \citet{2017SoPh..292..111R} illustrated that $m=2$ already exists in weak nonlinearity. However, recent simulations by \citet{2025A&A...693A.201S} revealed that the $m\geq2$ modes are attenuated by KHI and Raleigh-Taylor instability in highly nonlinear loop oscillations. This highlights the need to characterise the nonlinear regime and define a \lq\lq safe zone\rq\rq\, where $m\geq2$ modes are negligible and KHI remains the dominant nonlinear effect.
Quantitatively, this safe zone could be parameterised in terms of $\frac{V_0L}{C_{\rm k}R}$ or initial displacement amplitude $\xi_0/R$, using thresholds below which higher-order mode power remains at least an order of magnitude smaller than that of $m=1$.
%\textcolor{red}{[Q: are $m\geq2$ modes must-have in larger amplitude? A safe zone where $m\geq2$ modes are not significant. But \citet{2017SoPh..292..111R} shows they already exist in weak nonlinearity.]}

%Can CoE oscillation be approximated by our model? CoM vs. CoE.
Another consideration is that the model is formulated for the CoM oscillation, while observational data reflect the CoE. Justifying its use for CoE requires further care. 
Since the contribution functions of different wavelengths peak at different temperatures, and temperature correlates with density, it is reasonable to expect that each channel samples a different part of the loop, characterised by a distinct density threshold $\rho_T$. Empirically, the 171\,\AA\ channel primarily captures the dense loop core (i.e., $\rho_T$ close to $\rho_i$), whereas hotter channels such as 193\,\AA\ are more sensitive to the loop boundary (i.e., $\rho_T$ approaching $\rho_e$).
In our study, CoE oscillations in 171\,\AA\ match CoM displacements at a high $\rho_T$, while those in 193\,\AA\ correspond to lower $\rho_T$, with all other parameters consistent with the simulations. These empirical agreements support the use of our model to approximate CoE behaviour in certain conditions. This is reasonable because the model assumes a coherent oscillation of plasma above a certain density threshold, which aligns with the regions that contribute most to the emission at a given wavelength. 
This holds at an early oscillation stage, when turbulence is not yet well developed, the oscillation of CoE$_{\lambda}$ nearly matches that of CoM$_{\rho\geq\rho_{T,\lambda}}$. As the turbulent layer grows to a certain size, however, the CoE$_{\lambda}$ displacement becomes slightly smaller than that of CoM$_{\rho\geq\rho_{T,\lambda}}$, even exhibiting a phase shift. To reproduce this enhanced decay at the later stage while keeping the early CoM oscillation unchanged, the turbulence damping model of CoM requires a larger $\zeta$ and a smaller $\rho_T$. %\textcolor{red}{(because they are degenerate).}
Speaking of fitting the observations, classical linear models, such as the exponentially damped sine function, have also been shown to approximate the observed oscillations reasonably well within observational uncertainty \citep{2012A&A...539A..37P,2019ApJS..241...31N}. Therefore, selecting the most appropriate model requires rigorous comparison. One approach is Bayesian model comparison, as demonstrated in \citetalias{2025ApJ...991..208Z}, to assess whether our turbulence-damping function offers a better fit than classic linear models. If validated, the model can then be used for parameter inference, such as estimating the oscillation period and initial velocity. Other fitted parameters should be treated carefully, as they are coupled to determine the decay, and $\zeta$ may be overestimated.
While inversely estimated values of $\eta$ (or equivalently $\rho_T$) are affected by a combination of temperature, density, and loop deformation, they still offer a qualitative probe of the physical conditions. For example, low $\eta$ (or high $\rho_T$) in both 171\,\AA\ and 193\,\AA\ suggests a very hot external environment and high internal density.
In addition, a faster apparent decay in 193\,\AA\ oscillation than 171\,\AA, i.e., $\rho_{T,171}>\rho_{T,193}$, may indicate the oscillating loop is surrounded by a hotter ambient medium. For a cooler external temperature, the decaying kink oscillations seen in 171\,\AA\ may decay faster than those in hotter channels.

While the link between the contribution function of the imaging channel and $\rho_T$ inevitably leads to the wavelength-dependent apparent damping, other physical processes not considered here may also cause a similar damping pattern. It is well-known that non-adiabatic effects \citep{2019FrASS...6...57S} like thermal conduction, radiative losses, and thermal misbalance \citep{2017ApJ...849...62N, 2021A&A...646A.155D} lead to temperature-dependent damping for slow magnetoacoustic waves. They may come into play in kink waves as well. For example, loop cooling in time can lead to strong damping with a decrease in oscillation periods and an increase in spatial shift of the anti-node of the first harmonic from the apex in time \citep{2009ApJ...707..750M}. In this case, the resulting damping profile is clearly different from that predicted by our turbulence damping model, which has an increasing period from the initial values. However, a similar analysis, but considering the flow caused by the density decrease due to the cooling, illustrates that the wave amplitude grows by cooling \citep{2011SoPh..271...41R}. Anyway, caution is needed when interpreting the observed wavelength-dependent damping, as many factors could come into play and disentangling other thermodynamic processes from the turbulence is non-trivial.

%\textcolor{red}{To approximate the CoE oscillation based on mixing theory, we need to figure out the corresponding intensity weight on the density. For simplification, we take the weight as the response $G(n_e,T)$. Since the response at 171\,\AA generally synchronises with the density profile of the mixing layer, the weight can be roughly represented by the density itself, i.e., a higher density is associated with higher emissivity in 171\,\AA. 171 captures the core, which is confined by a higher $\rho_T$.}

A further limitation is that the model only approximates dynamics during the turbulence development phase. In reality, observed time series may contain data from the fully turbulent regime, where the model no longer applies. Signatures of turbulence saturation include abrupt frequency drifts and rapid amplitude decay, which could help identify when to truncate the data before applying the nonlinear turbulence damping function. Additionally, simulations show that once the kinetic energy of the turbulent layer approaches that of the loop core, turbulence develops more rapidly than expected, reaching a fully turbulent state earlier than predicted by KHI mixing theory, which assumes this transition occurs when the core’s momentum drops to zero. 
Including additional signals in the fully turbulent stage in fitting would change the posterior distribution \citep{Farrell_Lewandowsky_2018}, as the inference is driven by data that violate the assumptions of the model, reducing both accuracy and interpretability of the estimated parameters. 
For example, in the case of $\zeta = 5$, $T_i/T_e = 0.3,V_0=0.15)$, fitting the full signal gives $\rho_{T,171} = 1$, $\rho_{T,193} = 1$. However, truncating the signal to the first two cycles yields much more realistic values: $\rho_{T,171} = 2.89$, $\rho_{T,193} = 2.2$. Fits with the full time series give an underestimated $\rho_T$, possibly along with an overestimated $\zeta$. In simulation-based studies, truncation timing can be determined when the core kinetic energy equals that of the mixing layer. In observations, however, such estimation is nontrivial. A rough estimation is possible by identifying the turbulence-saturation signature mentioned above.
Another challenge is that if the oscillation duration is too short, say only one cycle, the CoM displacements above different $\rho_T$ thresholds and CoE oscillations across wavelengths may appear nearly identical with observational noise, making it difficult to accurately constrain $\rho_T$.
%because the mixing layer is not well developed.

%Why fit paired-oscillations rather than a single series? Gather more information.
%oscillations in one wavelength are independent of other wavelengths.

KHI-induced turbulence is a common feature in many numerical studies of transverse waves in loops \citep[e.g.,][]{2019ApJ...870...55G,2021ApJ...908..233S,2024A&A...688A..80K,2024A&A...689A.195G}, yet its effect on damping was only recently described analytically by \citetalias{2024ApJ...966...68H}. Once KHI presents, as evidenced by the small-scale vortices at the loop boundaries, we would predict, based on the result of \citetalias{2024ApJ...966...68H,2025ApJ...991..208Z}, that the resulting momentum exchange between the high-density core and its surroundings leads to wave damping. This is supported by \citet{2024A&A...688A..80K}, where amplitude decays as the turbulent layer thickens. Similarly, \citet{2015A&A...582A.117M} utilised a setup comparable to ours (but including radiation) and reported KHI structures (Fig.~6) and a damping profile (Fig.~9) that align with our model. Their comparison between the low and high-amplitude driving cases demonstrates that KHI turbulence induces strong damping beyond the radiative losses.
While the wave damping is common when KHI presents, where the energy eventually goes is highly dependent on the drivers. In some cases with continuous drivers, such turbulence cascades can generate sufficient heat to balance the radiative loss \citep[e.g.,][]{2021ApJ...908..233S}.

\section{Conclusion}\label{sec:conclusion}
In this study, we investigated the observational signature of nonlinearity, including KHI-induced turbulence, in large-amplitude impulsive driven kink oscillations of a straight coronal loop, using 3D MHD numerical simulations and forward modelling. The nonlinearity is quantified by a parameter $\frac{V_iL}{C_{\rm k}R}$. For $\frac{V_iL}{C_{\rm k}R}\geq1$, KHI-induced turbulence develops at the loop boundary and penetrates the core, damping the oscillation. This leads to frequency drift and time-varying damping rate in CoM displacement. An analytical model formulates this process.
In addition, higher-order modes are excited, causing cross-sectional compression and expansion, which enhances the energy cascade and damping, and reduces the oscillation amplitude relative to linear expectations. 
High-resolution 3D MHD simulations also reveal that the base period is longer than the linear kink period by a few percent, with the increase depending on density contrast and oscillation amplitude.
The physical reason for the period increase remained unexplained and will be investigated in a further study. %In observations, a few percent increase might be blurred by noise. 

Fitting the simulated oscillations with the turbulence-damping model using a Bayesian approach identifies significant parameter degeneracy within the model. While the primary observables, the initial velocity amplitude $V_i$ and effective oscillation period $\rm{P_k}$, are tightly constrained and reliably recovered, the parameters describing the damping profile (the mixing efficiency $C_1$, density contrast $\zeta$, and mixing-layer participation $\rho_T$) show correlated posteriors. This correlation leads to biased fitted values despite good overall agreement with the simulated signals. Therefore, in seismological applications with our model, $V_i$ and $\rm{P_k}$ can be meaningfully inferred from observations. However, other degenerate parameters require additional observational constraints to ensure confident inference.

Forward modelling shows that high-order modes and turbulence manifest as time-varying loop width and fine-scale, thread-like structures in synthetic EUV images. CoE oscillations retain key nonlinear features, including the base period increase, amplitude reduction, frequency drift, and time-varying damping. Oscillation patterns vary with loop segment, LoS, and wavelength. Different LoS angles introduce amplitude and slight phase shifts. Spatial resolution is critical. Features like loop width variation, LoS-dependent phase shifts, and small-scale structures are significantly blurred by noise at the AIA resolution. 
Multi-wavelength comparisons reveal phase differences and amplitude variations due to channel-specific temperature sensitivity.
In 171\,\AA, the oscillating loop exhibits decreasing brightness and thinner loop width in time, while 193\,\AA\ see loop brightness enhanced and the loop becomes broader. It is also found that CoE oscillations in hotter channels decay slightly faster with lower displacement and greater phase shift, compared to 171\,\AA\ oscillations. Similar behaviour is seen in previous studies with a similar setup and even in the case of uniform temperature \citep{2017ApJ...836..219A}.

%\textcolor{red}{In observations, we can identify the nonlinear kink oscillations by their time-varying damping behaviour in a single channel, and further by oscillations in multiple channels showing different apparent damping rates and phase shifts.}

Comparison between CoM oscillation and CoE oscillations shows both consistency and discrepancies. Generally, CoE oscillations in the 171\,\AA\ channel can be approximated by the CoM above a high $\rho_T$, while the 193\,\AA\ and 211\,\AA\ channels reflect lower-density plasma closer to $\rho_e$. This relationship holds for loops embedded in a hot background or under conditions of uniform temperature, but should reverse for hot loops surrounded by cooler plasma.
EUV CoE oscillations cannot be explained by temperature-density contribution functions alone. Early on, the CoE position matches the CoM of plasma above a temperature-derived density threshold, later divergence arises due to loop deformation. Turbulence and higher-order modes deform the loop’s cross-section, concentrating cooler emission in the core and shifting hotter emission toward the tail. This causes CoE displacements, especially in 193\,\AA\ and 211\,\AA, to systematically underestimate the motions. Observationally, this suggests that CoE signals in EUV channels may underrepresent the actual dynamics of oscillating loops embedded in hotter ambient plasma. The divergence appears as CoE oscillations at the later stage has smaller displacements and greater phase shift than that of CoM. Consequently, fitting the CoE oscillations using our model yields greater $\zeta$ and smaller $\rho_T$.

To summarise, the simulated impulsively driven nonlinear decaying transverse oscillations of coronal loops broadly agree with the KHI-induced turbulence damping model, despite the base period increase and excitation of $m\geq2$-modes in strong nonlinearity. Forward modelling the simulated oscillating loop in synthetic EUV observations preserves the time-varying damping behaviour in each channel, and oscillations in multiple channels showing different apparent damping rates and phase shifts.
The reported quantitative features enable the identification of nonlinear kink oscillations damped by KHI-induced turbulence in coronal loops. Remaining unexplained characteristics and model limitations motivate future investigation and model refinement. Follow-up research will improve the nonlinear damping models, quantify the role of higher-order modes, and employ multi-wavelength observations to identify nonlinear oscillations and validate emission bias in decaying kink oscillations.

\begin{acknowledgments}
\textbf{Acknowledgments}\\
We thank the anonymous reviewer for their constructive comments that helped to improved the clarity and quality of this work. We thank Prof.~Roberto Soler for inspiring discussions on the period increase in nonlinear kink waves. This work is supported by STFC Research grant No.~ST/Y00230X/1, project PID2024-156538NB-I00 from MCIN/AEI/10.13039/501100011033 and “ERDF A way of making Europe”. S.Z. acknowledges support by an FWO (Fonds voor Wetenschappelijk Onderzoek -- Vlaanderen) postdoctoral fellowship (1203225N).
For the purpose of open access, the author has applied a Creative Commons Attribution (CC BY) licence to any author-accepted manuscript version arising.\\
\textbf{Data Availability}\\
\noindent The FoMo code is available at \url{https://wiki.esat.kuleuven.be/FoMo}. Our version of forward modelling procedure/script written in Python, and an example notebook are available at \url{10.5281/zenodo.18258480} and \url{https://github.com/Sihui-Zhong/FoMo_Py}. The active branch of the (P\underline{I}P) code is available at \url{https://github.com/AstroSnow/PIP}. Simulation data are available upon reasonable request.
\end{acknowledgments}

\vspace{5mm}
%\facilities{AIA/SDO}


\software{FoMo \citep{2016FrASS...3....4V},
          PIPpy (\url{https://github.com/AstroSnow/PIPpy}),
          SciPy \citep{2020NatMe..17..261V}, 
          SSWIDL \citep{1998SoPh..182..497F},
          SunPy \citep{2020ApJ...890...68S},
          }


\bibliography{references}{}
\bibliographystyle{aasjournal}



\appendix
\section{The linear growth of the mixing layer}\label{append:1}
\begin{figure}[ht]
	\centering
    \includegraphics[width=0.6\linewidth]{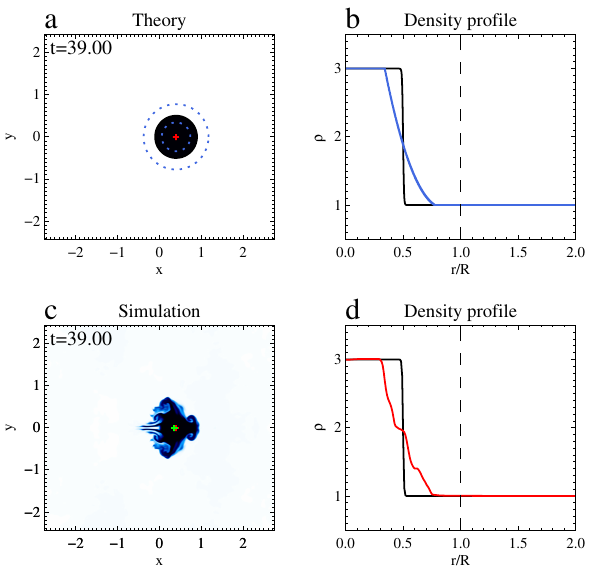}
	\caption{ 
        Comparison between the theoretical prediction (a--b) and simulation results (c--d) of the loop cross-section (a, c) and the transverse density profile (b, d) for the case of $\zeta = 3, V_0=0.15$. In panel (a), the bounds of the mixing layer are indicated by dashed circles. The plus symbols in panels (a) and (c) mark the loop centre. In panels (b) and (d), the black curve shows the initial density profile, the blue curve in panel (b) represents the analytically estimated average density profile, and the red curve in panel (d) shows the profile averaged over the azimuthal direction in the simulation. An animation illustrating the time evolution of the loop cross-section and transverse density profile is available. The real-time duration of the animation is 9~s. The animation runs from t = 0 to 104.5 s.
    }
	\label{fig:profile}
\end{figure}

According to \citetalias{2024ApJ...966...68H}, the KHI-induced mixing layer evolves self-similarly, with its thickness increasing linearly with time $t$ (see also Eq.~1 in \citetalias{2025ApJ...991..208Z}). This behaviour is confirmed by our 3D MHD simulations, where the azimuthally averaged density profiles agree well with the theoretical prediction. Figure~\ref{fig:profile} illustrates this for $\zeta=3$, along with an animation showing the evolution of the loop cross-section and density profile. Despite the loop's significant deformation from its initial circular shape, the averaged density profile (red in panel d) remains consistent with the analytical prediction (blue in panel b).

\section{Synthetic oscillations with FoMo}\label{append:B}
\begin{figure}[ht]
	\centering
	\includegraphics[width=\linewidth]{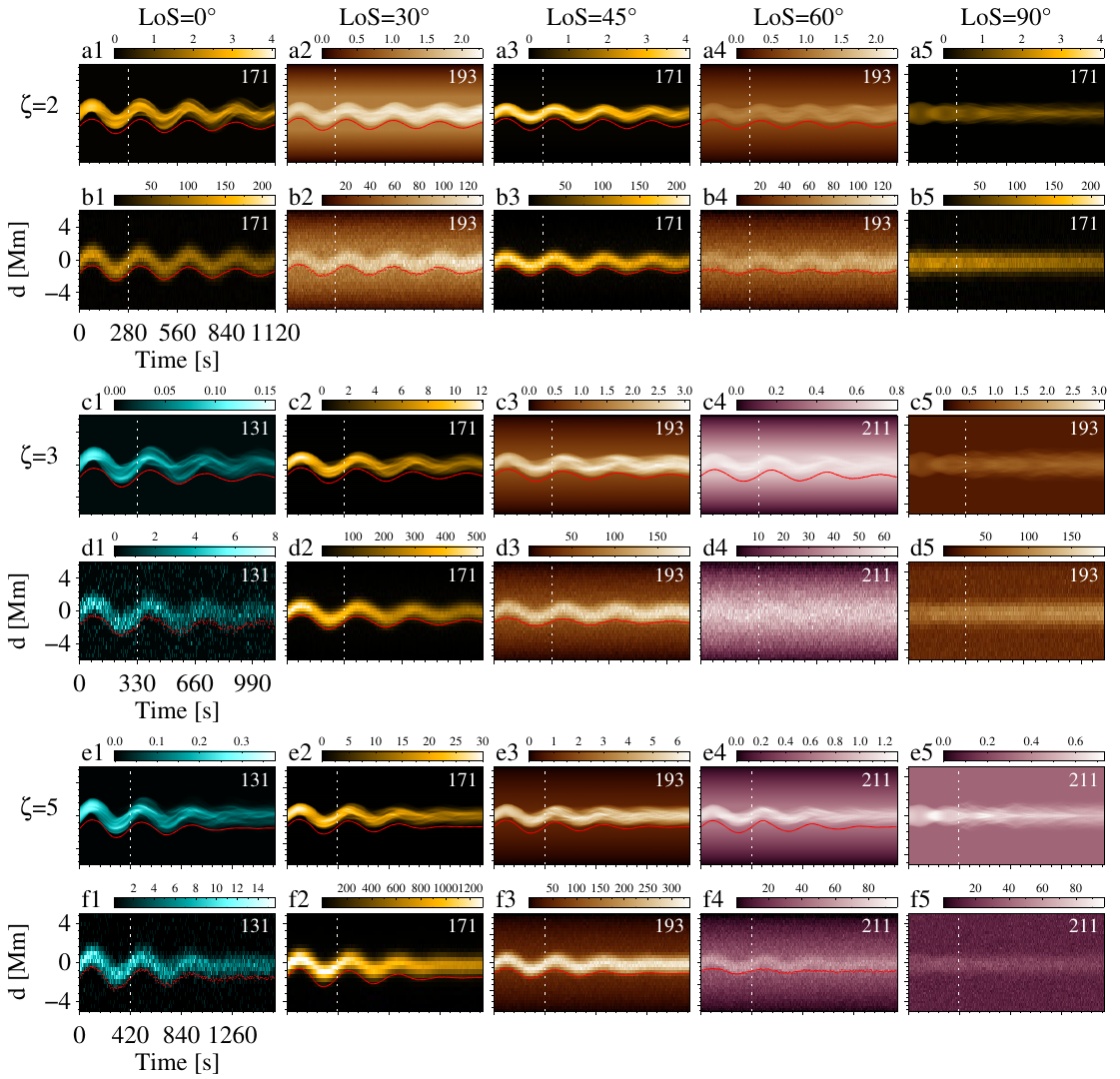}
	\caption{
        Time--distance maps for oscillating loops observed in different AIA channels and different LoS angles: $\theta=0$ (first column), $30^{\circ}$ (second column), $45^{\circ}$ (third column), $60^{\circ}$ (fourth column), $90^{\circ}$ (fifth column). The first, second and bottom two rows correspond to the loop with $\zeta=2,3,5$ and $T_i/T_e=0.5$. The displacement of the loop centre, tracked by Gaussian fitting, is indicated by red dots. The signals are shifted down by 1.6~Mm to avoid blocking the loop information. The colour bar is for intensity. %ADD time series of loop width? 
	}
	\label{fig:td}
\end{figure}

This subsection presents the oscillatory behaviour of different loop segments across multiple EUV channels and LoS angles.
Time--distance maps were made from synthetic images to reveal the oscillatory pattern observed in LoS angles spanning from $0^{\circ}$ to $90^{\circ}$. Loop centre positions were determined by Gaussian fits to the transverse intensity profiles, composing the oscillatory signals (red dots in Fig.~\ref{fig:td}). 
Fig.~\ref{fig:td} shows the time--distance maps at the loop apex for various $\zeta$, five LoS angles, and four AIA channels, at both original and AIA resolution.
In general, transverse loop oscillations are most clearly seen in the $0^{\circ}$ LoS. 
In $90^{\circ}$ LoS, parallel to the displacement direction, only loop compression and expansion are visible, without lateral motion.
Moreover, loop dynamics vary with wavelength. For $\zeta=2$, the 131\,\AA\ and 211\,\AA\ emissions are too faint to be detected. In 171\,\AA, loop brightness peaks at the first crest (e.g., Fig.~\ref{fig:td}e2, f2) and fades later, accompanied by a narrower apparent width. In 193\,\AA, the loop brightens and broadens in the later stage where turbulence penetrates deeper into the core, see examples in Fig.~\ref{fig:td}a2--b2 and c3--d3. The behaviour in 131\,\AA\ resembles that in 171\,\AA\ but with an intensity an order of magnitude smaller. The behaviour in 211\,\AA\ lies between that in 171\,\AA\ and that in 193\,\AA, with finer threads revealed at a later time.
Furthermore, thread-like fine structures are visible in all four channels at the original resolution, consistent with KHI signatures reported in earlier simulations \citep[e.g.,][]{2014ApJ...787L..22A,2017ApJ...836..219A}.
At AIA resolution, temporal variation in brightness and width are preserved, but those fine structures vanish, indicating that while AIA can capture large-scale loop deformation due to nonlinearity, it cannot resolve the turbulent scales.

\begin{figure}[ht]
	\centering
	\includegraphics[width=0.9\linewidth]{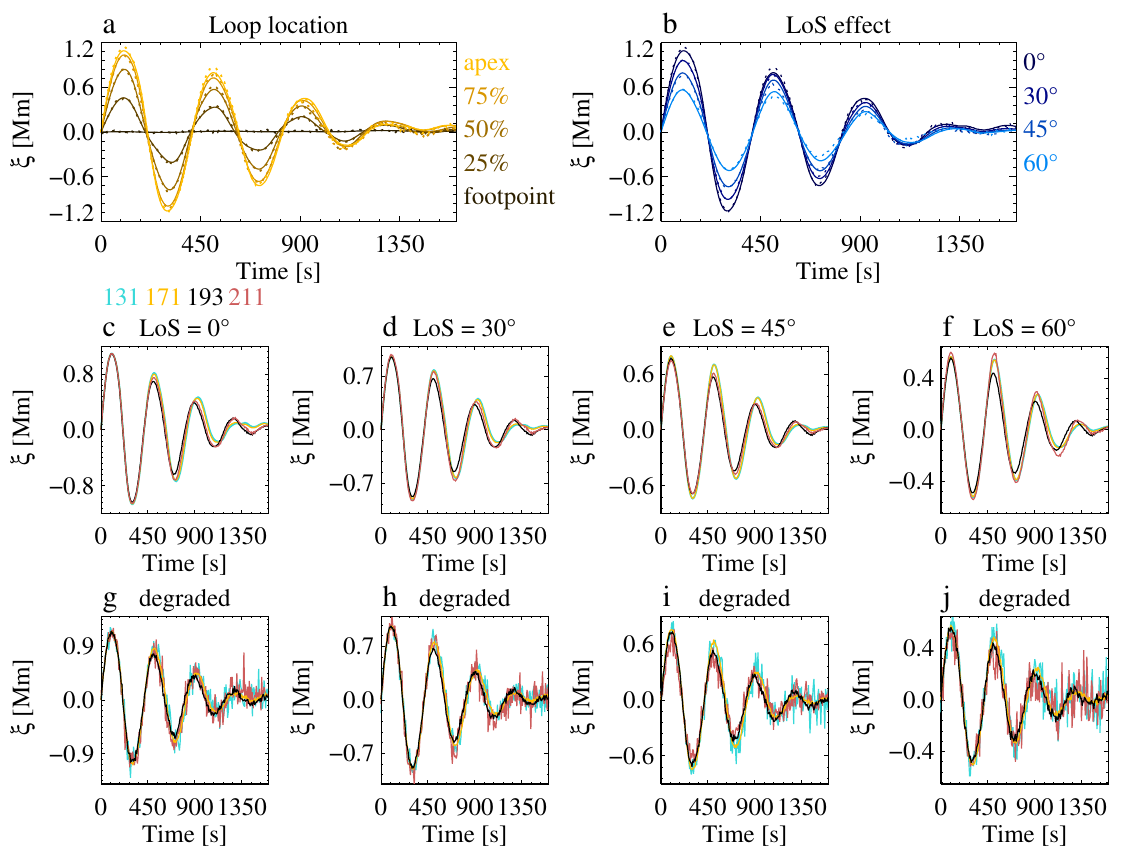}
	\caption{
        Oscillation signals of the loop with $\zeta=5, V_0=0.10$ extracted in different loop locations (a), LoS angles (b), and wavelengths (c--j). In panels (a--b), the signals are observed in 171\,\AA. The solid and dotted curves indicate the signals in original resolution and AIA resolution, respectively. Panels (c--f): oscillations detected in five LoS angles and four wavelengths with original simulation resolution, while the degraded versions are displayed in panels (g--j). %Loop radius = 1 Mm.
	}
	\label{fig:loop4_signals}
\end{figure}

A detailed comparison of oscillations across different loop segments, LoS angles, and wavelengths, is presented in Fig.~\ref{fig:loop4_signals} for a representative case with $\zeta=5, T_i/T_e=0.5$ and $V_0=0.10$. To study the impact of observational resolution, comparisons between the original and degraded AIA resolution are shown.
Oscillations at different loop locations share a similar pattern, but the amplitude decreases from the apex to the footpoint, where it vanishes, see solid curves Fig.~\ref{fig:loop4_signals}a. This indicates that the disturbance excites a standing mode. This standing-mode behaviour remains at degraded resolution (see dotted curves). 
Fig.~\ref{fig:loop4_signals}b shows that oscillations at different LoS angles exhibit a minor phase shift, in both original and degraded resolution. In $60^{\circ}$ LoS, the oscillation shows no decay (and in some cases grows, see Fig.~\ref{fig:loop41_signals} in the Appendix~\ref{append:B}) until the second crest, which, unfortunately, is lost when degraded to AIA resolution (see comparison between Fig.~\ref{fig:loop4_signals}f and j). This is a projection effect.

%\section{oscillations with higher external temperature}\label{append:B}
Fig.~\ref{fig:loop41_signals} shows synthetic EUV oscillations of the loop with $\rho_i=5,T_i/T_e=0.3,V_0=0.10$. Compared to oscillations in the loop with lower external temperature (Fig~\ref{fig:loop4_signals}), oscillations in 171\,\AA\ and 193/211\,\AA\ exhibit greater differences in apparent damping rate, including both greater amplitude drops and larger phase shifts in the latter.

\begin{figure}[ht]
	\centering
	\includegraphics[width=\linewidth]{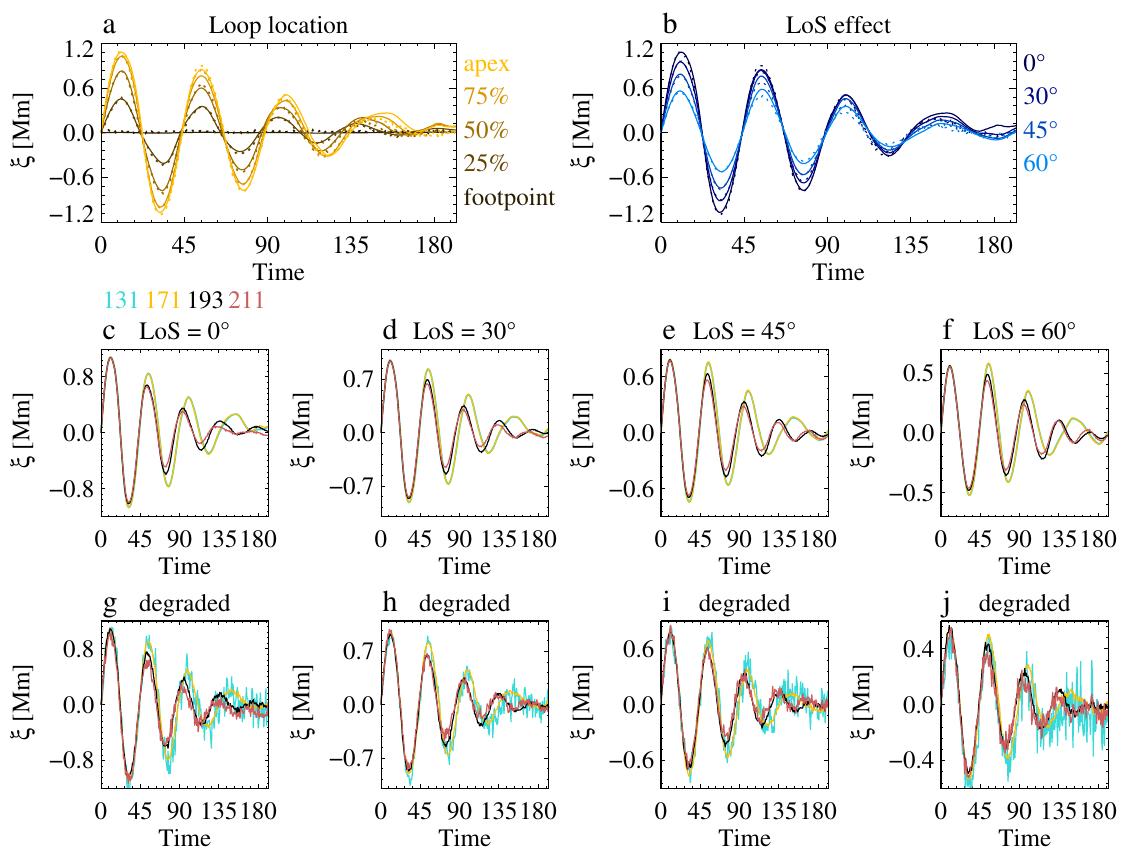}
	\caption{
       Similar to Fig.~\ref{fig:loop4_signals} ($\rho_i=5$) but with $T_i/T_e=0.3$. %Zhong\_Loop41.
	}
	\label{fig:loop41_signals}
\end{figure}

\end{document}